%% file: SOFI.tex
\documentclass{aa}
\usepackage{amsmath,txfonts,graphicx,natbib,epsfig,amssymb}
\usepackage[figuresright]{rotating}
\bibpunct{(}{)}{;}{a}{}{,}

\input{rrmacros2e.tex}

\begin{document}

\title{Near-IR Spectroscopy of OH/IR stars in the Galactic Centre}
\author{E. Vanhollebeke \inst{1} \and J.A.D.L. Blommaert \inst{1} \and
  M. Schultheis \inst{2} \and B. Aringer \inst{3} \and A. Lan\c{c}on \inst{4}}
\institute{\Leuven \and CNRS UMR6091, Observatoire de Besan\c{c}on,
  BP1615, F-25010 Besan\c{c}on, France \and \Wien \and \Strasbourg }

\offprints{E. Vanhollebeke, \email{evelien@ster.kuleuven.be}}

\date{Received / Accepted }

\abstract
{Galactic Centre (GC) OH/IR stars can be, based on the expansion 
  velocities of their circumstellar shells, divided into two groups 
  which are kinematically different and therefore are believed to 
  have evolved from different stellar populations.}
{To study the metallicity distribution of the OH/IR stars population in 
the GC on basis of a theoretical relation between EW(Na), EW(Ca) 
and EW(CO) and the metallicity.}
{For 70 OH/IR stars in the GC, we obtained near-IR spectra. 
 The equivalent line widths of
\ion{Na}{i}, \ion{Ca}{i}, \element[][12]{CO}(2,0) and the curvature of
the spectrum around 1.6~\micron\ due to water absorption are determined.}
{The near-IR spectrum of OH/IR stars is influenced by several physical
  processes. OH/IR stars are variable stars suffering high
  mass-loss rates. The dust that is formed around the stars strongly
  influences the near-IR spectra and reduces the  equivalent line widths
  of \ion{Na}{i}, \ion{Ca}{i}. A similar effect is caused by the water 
  content in the outer atmosphere of the OH/IR star.
  Because of these effects, it is not possible with our low resolution 
  near-infrared spectroscopy to determine the metallicities of these stars.}
{}



\maketitle


\section{Introduction}
\label{section introduction}
The inner region of our Galaxy has been extensively searched for OH/IR stars
in the past two decades. The studies of the GC region are 
severly hampered, especially in the visible wavelength range, by extinction
and also source confusion. Searches for OH/IR stars at radio wavelengths at 
the 1612~MHz OH maser line were therefore very useful and also provided 
kinematic information of the detected stars. 
\citet{Lindqvist1992b} searched for OH/IR stars in six VLA 
primary beam fields and identified 134 OH/IR stars. Further searches, also 
using ATCA, were conducted by \citet{Sevenster1997}, who found 145 new OH
masering sources in the Bulge and \citet{Sjouwerman1998} who discovered 
an additional 52 new OH/IR stars in the GC region. A review on 
Asymptotic Giant Branch (AGB) stars in the GC region can
be found in \citet{HabingWhitelock2003}.\\
\citet{Lindqvist1992a} studied the spatial and
kinematic properties of their OH/IR sample and divided the
stars into two groups on basis of their outflow velocities. 
OH/IR stars with small expansion velocities 
($v_{\rm exp} < 18.0$~km/s) 
have a larger spread in latitude and
a larger velocity dispersion with respect to the galactic rotation
than the group with higher expansion velocities. This was also found by  
\citet{Baud1981} for OH/IR stars in the galactic disk.
The low expansion velocity stars 
are expected to be older objects with larger peculiar
motions, whereas the other group may have a different formation history;
it might be a later addition to the GC, possibly via a merger.  
The outflow velocity of the circumstellar shell 
is related to the luminosity of the star and to the properties of the dust in
the circumstellar shell \citep{Tignon1994,Elitzur2003}. \\
Several groups have been searching for the infrared counterparts of the OH
maser sources \citep{Jones1994,Blommaert1998,Wood1998,Ortiz2002}.  
It would be expected that the 'older' stars have lower luminosities
than the high expansion group. \citet{Blommaert1998} found that the
high expansion velocities group contains higher luminosity stars but that
there was also a large overlap in the luminosity distributions of the
2 groups.  \citet{Ortiz2002} also compared the measured luminosities
using a larger number of stars and did not find any evidence for a
distinction on basis of luminosities between the 2 groups. The
disadvantage of the latter study is the fact that the luminosities
were not corrected for variability and that the luminosities spread
over a larger range.  Nevertheless, it seems that the differences in
luminosities between the 2 groups, if they exist, are not very strong.
If the luminosity is not sufficient to explain the differences in
expansion velocities than it would be expected that differences in the
gas to dust ratios and thus metallicities of the stars must exist. It
was demonstrated by \citet{Wood1992} for OH/IR stars 
in the Large Magellanic Cloud and by \citet{Blommaert1993} for OH/IR
stars in the 
outer Galaxy that the expansion velocities are low, even though the
OH/IR stars have high luminosities. In both, the LMC and in the outer Galaxy,
it is expected that the stars have indeed low metallicities. The next
logical step would be to investigate the metallicities of GC OH/IR stars.\\
\citet{Schultheis2003} obtained near-IR spectra of 107 sources with
mid-infrared excess selected from the ISOGAL survey \citep{Omont2003},
including 15 OH/IR sources. With an empirical formula based on near
infrared spectroscopy of K and M giants \citep{Ramirez2000,Frogel2001},
\citet{Schultheis2003} tentatively estimated [Fe/H] for all of the ISOGAL
sources. Although \citet{Schultheis2003} indicate that the 
\citet{Ramirez2000} formula does not actually measure metallicity for
an individual spectrum of a strongly variable star, it was found that
it might still be used to find an average metallicity.\\

In this paper we will discuss our attempt to apply the results of
\citet{Schultheis2003}  on medium-resolution
near-IR spectra on a sample of GC OH/IR stars. 
Three things will be discussed: we will take a look
at the influence of the water content on the atomic lines \ion{Na}{i}
and \ion{Ca}{i}, we discuss how periodicity can have an influence on
the water content and therefore also influences the near-IR spectrum
and we will study a grid of dust models and show that hot dust also
has an influence on the near-IR spectrum.\\
 
The sample, observations and data-reduction will be described in the
next section. In section \ref{section analysis}, we explain our method
of analyse and compare spectra we have in common with other
authors. The problems one encounters analysing near-IR spectra of OH/IR
stars will be discussed in section \ref{section discussion}. Finally
in section \ref{section conclusions} we summarise and come to the
conclusions.\\


\section{Observations and data reduction}
\label{section observations and data reduction}

\subsection{Sample}
\label{section sample}
The sample consists of 70 OH/IR stars located in the GC region. Almost all
stars are selected from \citet{Lindqvist1992b} and
\citet{Sjouwerman1998}. Our sample also includes the 15 OH/IR stars 
observed by \citet{Schultheis2003} so that we can compare our results and 
investigate the effect of the variability of this type of stars. 
We also selected the three "high-velocity" OH/IR stars that were detected 
in the direction of the GC \citep{vanLangevelde1992}. 
As we wanted to apply the metallicity versus EW(Na), EW(Ca) and EW(CO) 
calibration used in \citet{Ramirez2000} and \citet{Frogel2001}
to our sample we also observed 11 stars from
\citet{Ramirez1997} (K-band spectra of 43
luminosity class III stars from K0 to M6) and \citet{Ramirez2000} (K-band
spectra of more than 110 M giants in the Galactic Bulge (GB)). 
Finally, to investigate the effects of the variability of our Long Period
Variable stars (LPVs) on our analysis we measured variable stars from 
\citet{Lancon2000} (spectra of cool, mostly variable, giant and
supergiant stars) (see Table~\ref{table sample 28/06} -- Table~\ref{table
sample 30/06}).

\subsection{Observations}
\label{section observations}
The near-IR spectra were obtained with the 3.58~m NTT (ESO) at la Silla, Chile
between June 28th - 30th 2003 using the red grism of the SOFI
spectrograph. This resulted in spectra from 1.53 \micron\ up to 2.52
\micron. Before the actual spectrum was obtained, the instrument was
used in imaging mode, to acquire the star in the slit. Several spectra
were taken with the star in different places along the slit. Standard
stars of spectral type O till G were observed as close as possible to
the object stars' airmass in order to correct for telluric absorption
features.

\subsection{Data reduction}
\label{section data reduction}
The data reduction was done using the ESO-Munich Image Data Analysis
System (ESO-MIDAS). The images were first corrected for cosmic ray
hits. Several spectra of the same target along the slit were obtained
and subtracted from each other to correct for the sky level. The
images were flat-fielded using dome flats. A Gaussian fit
perpendicular to the dispersion direction was used to subtract a one
dimensional spectrum out of the two dimensional image: columns that
fall within the FWHM of the Gaussian fit were added to the
spectrum (3-4 columns on average). A correction for distortion along
the slit was unnecessary. During the extraction process, a correction
for bad pixels was made, 
they were left out of the spectrum, no averaging was done around these
pixels. The wavelength calibration was based on the spectrum of a
Xenon lamp. This image was also flat-fielded using the dome flats and
corrected for bad pixels in the same way as the other spectra.  The
wavelength calibration resulted in a dispersion of typically 10.13
\AA/pixel and $\lambda/ \Delta \lambda \approx 1000$.\\ 
The standard stars
were reduced in the same way. Thereafter they were divided by a
Kurucz-model that corresponds to their spectral type. The resulting
curve was used to correct the objects for telluric lines and also to
correct for the instrumental response function. Different curves for
different airmasses were made in order to correct for the airmass.\\
The dereddening law used, is based on \citet{Cardelli1989}, $A_{\rm v}$
values came from \citet{Schultheis1999}.\\ The resulting spectra are
shown in Fig.~\ref{figure OH/IR 1} till Fig.~\ref{figure Lancon and Wood}.\\ 


\section{Analysis}
\label{section analysis}

\begin{table}
  \caption{Definition of band passes for continuum
    and features \citep{Schultheis2003}.}
  \label{table EW}
  \begin{center}
    \begin{tabular}{lc}\hline\hline
      Feature & Band passes [\micron]\\ \hline \ion{Na}{i} feature & 2.204
      -- 2.211\\ \ion{Na}{i} continuum \#1 & 2.191 -- 2.197\\ \ion{Na}{i}
      continuum \#2 & 2.213 -- 2.217\\ \ion{Ca}{i} feature & 2.258 --
      2.269\\ \ion{Ca}{i} continuum \#1 & 2.245 -- 2.256\\ \ion{Ca}{i} continuum
      \#2 & 2.270 -- 2.272\\ \element[][12]{CO}(2,0) feature band head & 2.289 --
      2.302\\ \element[][12]{CO}(2,0) continuum \#1 & 2.252 -- 2.258\\ \element[][12]{CO}(2,0)
      continuum \#2 & 2.284 -- 2.291\\ \HtwoO\ continuum & 1.629 --
      1.720\\ \HtwoO\ absorption wing 1 & 1.515 -- 1.525\\ \HtwoO\
      absorption wing 2 & 1.770 -- 1.780\\\hline
    \end{tabular}
    \end{center}
\end{table}

\begin{figure}
  \resizebox{\hsize}{!}{\includegraphics{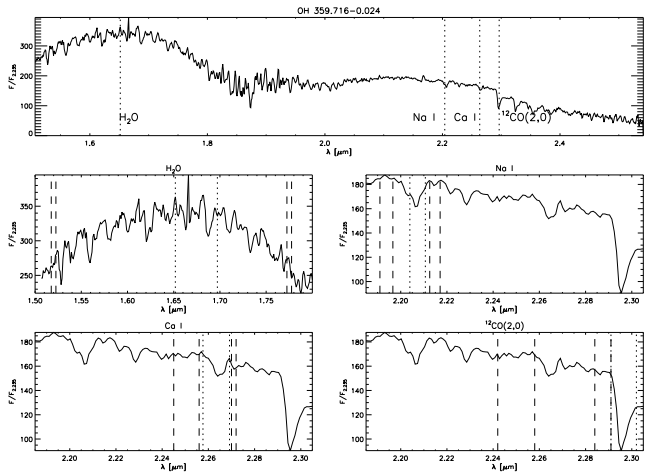}}
  \caption{An overview of the chosen definition of band passes for
  continuum and features. The top figure shows the overall spectrum
  for a typical OH/IR stars. The dotted lines indicate the central
  positions of the features. The other 4 figures show for each line
  the selected passband for the feature (dotted lines) and the
  selected passband for the continuum (dashed lines). (For the
  \element[][12]{CO}(2,0) the feature's first dotted line overlaps
  with the last dashed line from the continuum.)}
  \label{figure overzicht}
\end{figure}

\subsection{Equivalent line widths}
\label{section equivalent widths}
The equivalent line widths of \ion{Na}{i}, \ion{Ca}{i} and
\element[][12]{CO}(2,0) 
(see Tables~\ref{table sample 28/06}~--~\ref{table sample 30/06}) were
obtained in exact the same way as by \citet{Schultheis2003} \citep[see
also][]{Ramirez1997,Lancon2000}. They were measured relative to the
selected continuum bands using the Image Reduction and Analysis
Facility (IRAF) (see Table~\ref{table EW} and Fig.~\ref{figure
  overzicht}). \\ \citet{Ramirez1997} discuss how the equivalent width
measurements of \ion{Na}{i} and \ion{Ca}{i} in medium resolution
spectra contain contributions from other elements such as Sc, Ti, V,
Si and S. Thus, it turns out that a significant contribution of the
\ion{Na}{i} and \ion{Ca}{i} features is due to other species. In the
studied spectral region (1.53-2.52 \micron) the CN molecule causes a
noisy continuum \citep{Origlia1993}, introducing a pseudo-continuum
opacity (this is clearly shown in the high-resolution spectrum of RX
Boo (an M8 III star) in \citet{Wallace1996}). According to
\citet{Ramirez1997} the continuum bands that where used to determine
the EW(Ca) get affected by CN absorption for stars with \Teff$<
3000$~K. However, synthetic CN spectra based on hydrostatic
\textsc{Marcs} models for giant stars (for a description of such
models see \citet{Aringer1997}, in our case we assumed
$\log(G$~[cm/s$^2]) = 0.0$, solar mass and elemental abundances) show
that CN is important in all cool objects below 4000~K, and even gets
weaker below 3000~K.\\ The effective temperature has a strong impact
on the \ion{Ca}{i} and \ion{Na}{i} features: as \Teff\ decreases, the
equivalent widths of both features increase. Tables 
\ref{table linelist Na}, \ref{table linelist Ca} and \ref{table
  linelist CO} give an overview of what the respectively \ion{Na}{i},
\ion{Ca}{i}
and \element[][12]{CO}(2,0) lines in a medium resolution spectrum are
really made of. The tables are based on the high-resolution spectrum
of RX Boo in \citet{Wallace1996}. For cooler
oxygen--rich stars \citep{Lancon2000}, such as OH/IR stars, \HtwoO\
absorption, instead of CN absorption, becomes very important (see
section \ref{section water absorption in OH/IR stars}).\\

\subsection{Water absorption}
The amount of water has been obtained by measuring
the curvature in the spectrum around 1.6~\micron, as in
\citet{Schultheis2003}. The apparent bump around 1.6~\micron\ is caused
by strong and wide water absorption bands around 1.4 and
1.9~\micron. Formally, our measurements are equivalent widths of this
bump relative to reference fluxes measured on either side of it in
the wings of the water bands. As for the other 
equivalent line width 
measurements, its value is given in \AA, but it takes negative values
when water absorption is present (see Tables~\ref{table sample
28/06}~--~\ref{table sample 30/06}).

\subsection{Comparison with previous work}
\label{section comparison}

\begin{figure}
  \resizebox{\hsize}{!}{\includegraphics{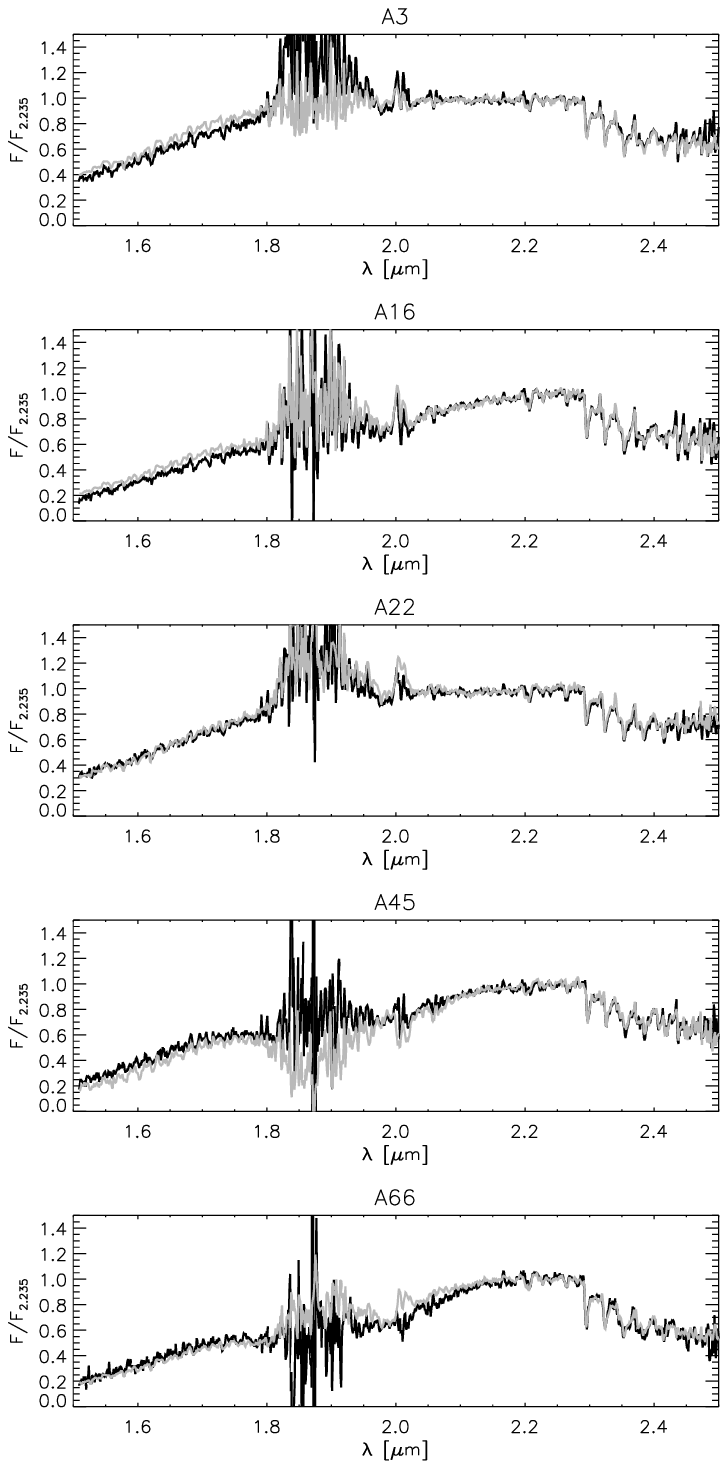}}
  \caption{RGB stars in common with \citet{Schultheis2003}; black
    line: this work, gray line: \citet{Schultheis2003}. The features
    between 1.8 and 1.9 \micron\ and around 2 \micron\ are due to the
    atmosphere.}
  \label{figure vergelijk Mathias volledig}
\end{figure}

\begin{figure}
  \resizebox{!}{\hsize}{\includegraphics{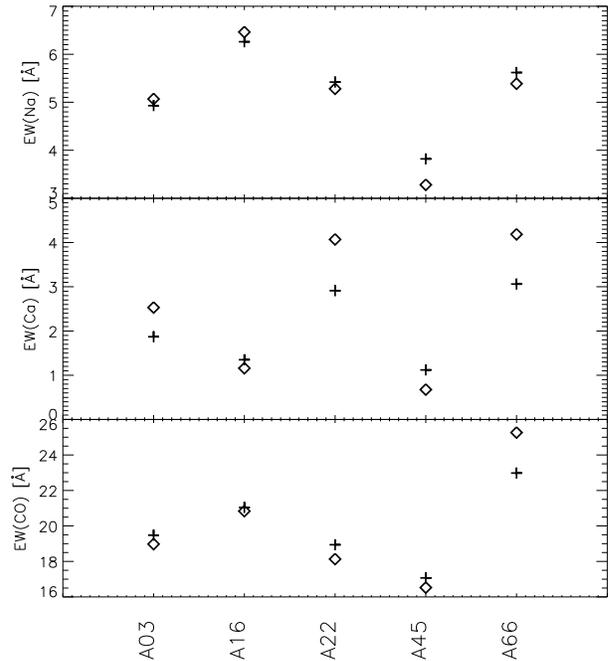}}
  \caption{Equivalent line width measurements for the candidate RGB
    stars in common 
    with \citet{Schultheis2003}. The plus-signs are taken from
  \citet{Schultheis2003}, the diamonds from this work.}
  \label{figure verschil EW Mathias}
\end{figure}

We have 5 candidate Red Giant Branch (RGB) stars in common with
\citet{Schultheis2003}. These candidate RGB stars, are likely
genuine ones, since we see no significant difference between
\citet{Schultheis2003} observations of these stars and our
observations. Stars on the RGB are expected to show weak variations
compared to OH/IR stars and Miras.\\ \citet{Schultheis2003} observed the
stars with the same instrument (SOFI), using the same grism. This
results in spectra with the same resolution as our spectra. The
time difference between these observations is about 3 years. The spectra
(normalised at 2.235 \micron) are compared in Fig.~\ref{figure
vergelijk Mathias volledig} and Table \ref{table comparison with
  Schultheis} gives the mean values and standard deviations for the
stars we have in common. The absolute average differences in equivalent
line widths for these RGB stars are consistent with scatter due to
formal errors ($\sim 1\AA$). One 
expects these errors to come from data-reduction issues, stability of
observing conditions during the night. The
absolute average differences are: 0.25 $\pm$ 0.17 \AA\ for EW(Na), 0.72 $\pm$
0.42 \AA\ for EW(Ca) and 0.87 $\pm$ 0.82 \AA\ for EW(CO) (see also
Fig.~\ref{figure verschil EW Mathias}).

\begin{figure}
  \begin{center}
    \resizebox{!}{\hsize}{\includegraphics{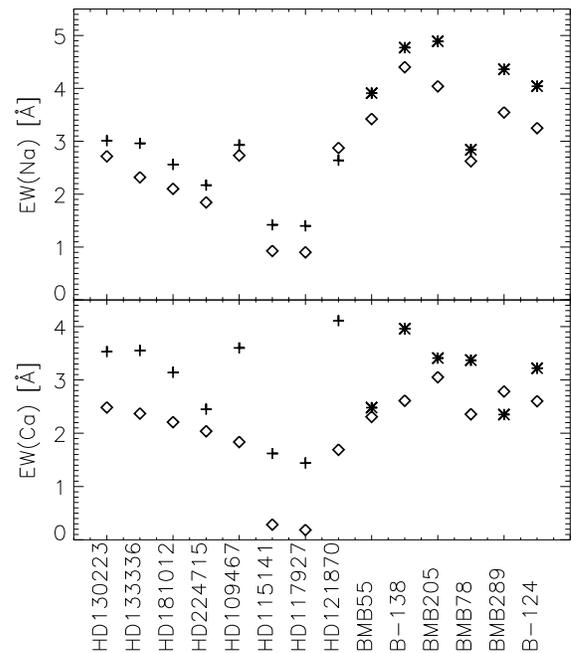}}
    \caption{Equivalent line width measurements for the stars in
      common with \citet{Ramirez1997} and \citet{Ramirez2000}. The
      plus-signs are taken from
      \citet{Ramirez1997}, the stars are from \citet{Ramirez2000} and
      the diamonds are from this work.}
    \label{figure verschil EW Ramirez}
  \end{center}
\end{figure}

We have 8 late-type giants in common with \citet{Ramirez1997} and
6 red giants situated in the Bulge with \citet{Ramirez2000}. A
comparison was made only for EW(Na) and EW(Ca), since
\citet{Ramirez1997} uses different continuum pass bands for obtaining
EW(CO). Tables \ref{table comparison with Ramirez1997} and \ref{table
  comparison with Ramirez2000} give the mean values for the compared
equivalent line widths. The absolute average differences with
\citet{Ramirez1997} are: 0.39 
$\pm$ 0.13 \AA\ for EW(Na) and 1.29 $\pm$ 0.41 \AA\ for EW(Ca). The
absolute average differences with \citet{Ramirez2000} are 0.59 $\pm$
0.23 \AA\ for EW(Na) and 0.66 $\pm$ 0.35 \AA\ for EW(Ca). The
difference in EW(Na) \citep{Ramirez1997,Ramirez2000} and EW(Ca)
\citep{Ramirez2000} is again consistent with scatter due to formal
errors. There is  no obvious reason why the difference for EW(Ca)
\citep{Ramirez1997} is larger than $\sim 
1\AA$ (see also Fig.~\ref{figure verschil EW Ramirez}), although we
have to keep in mind that there is a difference in resolution between
our spectra and the spectra of \citet{Ramirez1997,Ramirez2000}, which
can influence the difference between the measured equivalent line
widths. The data from \citet{Ramirez2000} can be downloaded from the
internet. After rebinning the data to our lower resolution, the
following values could be measured: $3.95 \pm 0.78$ for the EW(Na) and
$2.77 \pm 0.47$ for the EW(Ca), which resembles better the values we
found (see Table~\ref{table comparison with Ramirez1997}). A similar
test for \citet{Ramirez1997} could not be done.

\begin{figure}
\begin{center}
\resizebox{\hsize}{!}{\includegraphics{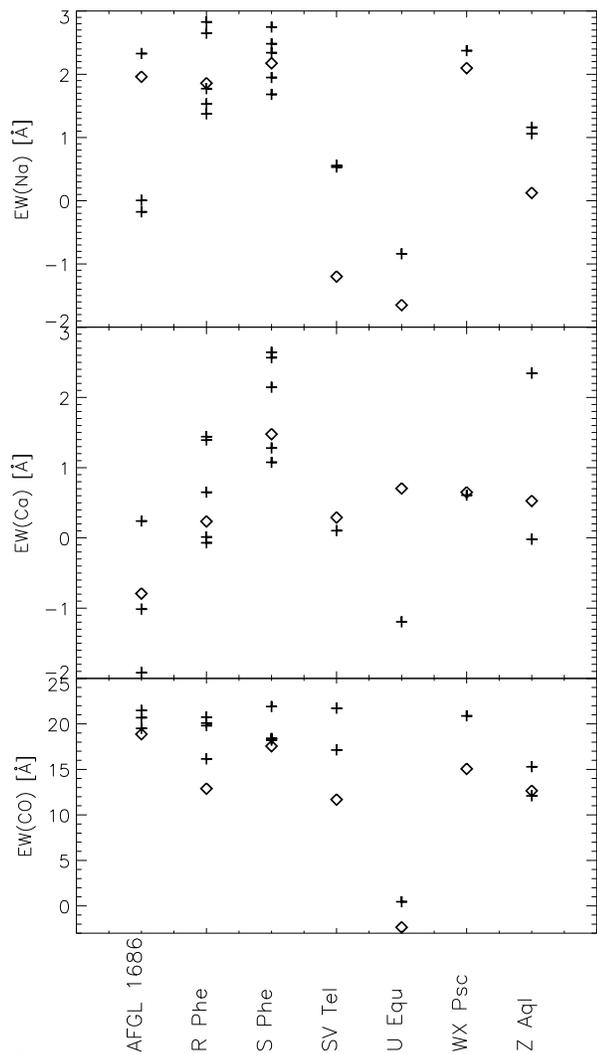}}
\caption{Variations in the equivalent widths for stars in common with
  \citet{Lancon2000}. The plus-signs are taken from
  \citet{Lancon2000}, the diamonds from this work.}
\label{figure verschil EW Lancon}
\end{center}
\end{figure}

\begin{figure}
\begin{center}
\resizebox{!}{\hsize}{\includegraphics{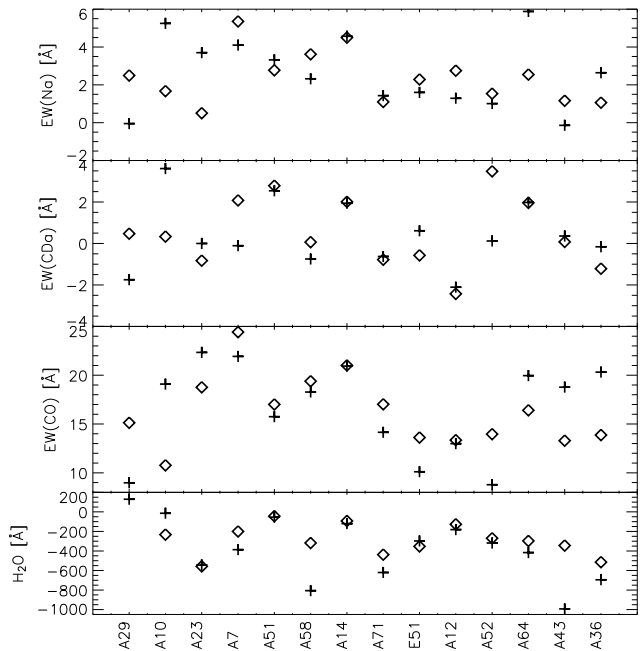}}
\caption{Equivalent line width measurements for the OH/IR stars in
  common with \citet{Schultheis2003}. The plus-signs are taken from
  \citet{Schultheis2003}, the diamonds are from this work. 
}
\label{figure vergelijking OH/IR Mathias}
\end{center}
\end{figure}

We have 7 luminous cool stars in common with \citet{Lancon2000}.
Three of these stars (WX Psc, AFGL 1686 and U Equ) are OH/IR stars
\citep{Lancon2000}. These OH/IR stars (not situated in the GC) are
used as a comparison for the equivalent line width measurements. The
other 4 stars, have periods 
in the range from 120 to 270 days. Some of the stars were observed
several times by \citet{Lancon2000}. The differences in equivalent
line widths between the measurements based on the available spectra of
\citet{Lancon2000} and our measurements are shown
in Fig.~\ref{figure verschil EW Lancon}. The differences seen in
this plot, are large in
comparison with the differences seen in Fig.~\ref{figure verschil
EW Mathias} and Fig.~\ref{figure verschil EW Ramirez}, especially 
for EW(CO),  and can't be
explained only by formal errors. The variability of the stars is
responsible for the large variations in the equivalent line widths
(see section \ref{section periodicity}).  Our measured equivalent
line widths do fall within the range of equivalent line widths based
on the spectra of \citet{Lancon2000} (see
Fig.~\ref{figure verschil EW Lancon}). We can expect that the OH/IR
stars that we have in common with \citet{Schultheis2003} will show
similar variations in their equivalent line widths, as is illustrated
in Fig.~\ref{figure vergelijking OH/IR Mathias}. The mean
absolute differences for the OH/IR stars in common with \citet{Schultheis2003}
concerning the equivalent line widths are: 1.56 $\pm$ 1.17 \AA\ for
EW(Na), 1.14 $\pm$ 1.16 \AA\ for EW(Ca), 3.60 $\pm$ 2.47 \AA\ for
EW(CO) and 171.81 $\pm$ 193.03 \AA\ for the water
absorption. These variations are caused by the variability of these
OH/IR stars, as will be explained in the following section.

\begin{table}
  \caption{Mean values for the equivalent line widths for the stars in
  common with \citet{Schultheis2003}.}
  \label{table comparison with Schultheis}
  \begin{center}
    \begin{tabular}{ccc}\hline\hline
      & this work & \citet{Schultheis2003} \\ \hline
      \ion{Na}{i} & $5.10 \pm 0.74$ & $5.21 \pm 0.67$ \\
      \ion{Ca}{i} & $2.52 \pm 1.29$ & $2.06 \pm 0.74$ \\
      \element[][12]{CO}(2,0) & $19.95 \pm 2.48$ & $19.90 \pm 1.69$ \\ \hline
    \end{tabular}
  \end{center}
\end{table}

\begin{table}
  \caption{Mean values for the equivalent line widths for the stars in
  common with \citet{Ramirez1997}.}
  \label{table comparison with Ramirez1997}
  \begin{center}
    \begin{tabular}{ccc}\hline\hline
      & this work & \citet{Ramirez1997} \\ \hline
      \ion{Na}{i} & $2.05 \pm 0.62$ & $2.39 \pm 0.54$ \\
      \ion{Ca}{i} & $1.64 \pm 0.70$ & $2.93 \pm 0.82$ \\ \hline
    \end{tabular}
  \end{center}
\end{table}

\begin{table}
  \caption{Mean values for the equivalent line widths for the stars in
  common with \citet{Ramirez2000}.}
  \label{table comparison with Ramirez2000}
  \begin{center}
    \begin{tabular}{cccc}\hline\hline
      & this work & \citet{Ramirez2000} & \citet{Ramirez2000}\\
      &           &                     & after rebinning\\ \hline
      \ion{Na}{i} & $3.55 \pm 0.45$ & $4.13 \pm 0.54$ & $3.95 \pm 0.78$\\
      \ion{Ca}{i} & $2.62 \pm 0.20$ & $3.13 \pm 0.48$ & $2.77 \pm 0.47$\\ \hline
    \end{tabular}
  \end{center}
\end{table}

\begin{figure}
\begin{center}
\resizebox{\hsize}{!}{\includegraphics{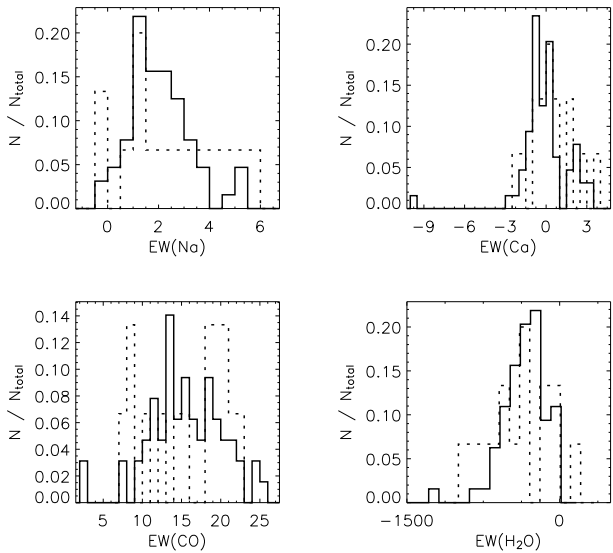}}
\caption{Histograms for EW(Ca), EW(Na), EW(CO) and the water
  absorption for the OH/IR stars from this work (full line) and the
  OH/IR stars from
  \citet{Schultheis2003} (dotted line). All x-axes are in the same unit \AA.}
\label{figure histoEW}
\end{center}
\end{figure}

\begin{figure}
\begin{center}
\resizebox{\hsize}{!}{\includegraphics{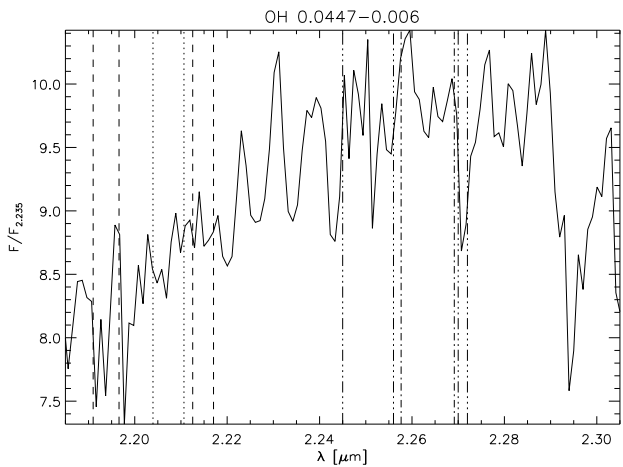}}
\caption{OH/IR star for which the \ion{Na}{i} and \ion{Ca}{i} lines
  couldn't be measured. The dotted lines indicate the \ion{Na}{i}
  feature and the dashed lines the selected continuum points. The
  dashed-dotted lines indicate the \ion{Ca}{i} feature with it's
  corresponding continuum (dashed-triple-dotted).}
\label{figure bad Ca Na}
\end{center}
\end{figure}

Fig.~\ref{figure histoEW} gives an overview of the equivalent line
widths for \ion{Na}{i}, \ion{Ca}{i} and \element[][12]{CO}(2,0) (see
section \ref{section equivalent widths} and Table~\ref{table EW}). The
fourth histogram gives also the amount of
water. Only the OH/IR stars are considered in this figure. Notice that
the EW(Na), EW(Ca), EW(CO) and the EW(\HtwoO) are all in the same unit
\AA. The EW(\HtwoO) are all negative values, indicating water absorption
in the spectrum around 1.6 \micron, as can be seen from the spectra of
the OH/IR stars in Figure \ref{figure OH/IR 1}.  A negative value for
EW(Ca) and EW(Na), indicates that the line couldn't been measured, we
do not expect these lines in emission. Fig~\ref{figure bad Ca Na}
gives an example of this. Especially for \ion{Ca}{i} it is clear why
the feature couldn't be measured: the second continuum band pass shows
a very deep feature, which causes the flux measurement in the feature
to be higher than in the continuum. This is also the case for the
\ion{Na}{i} feature, but here the first continuum band pass causes it
(less clear than for the \ion{Ca}{i} feature).\\

Fig. 2 in \citet{Ramirez2000} gives a 1 -- 5 \AA\ range for EW(Ca) and
a 2 - 6 \AA\ range for EW(Na) for a sample of red giants in the
GB. Fig.~\ref{figure histoEW} shows that 
our ranges for the equivalent widths for both \ion{Ca}{i} and
\ion{Na}{i} are lower. Histograms for the equivalent line widths of
\ion{Na}{i}, \ion{Ca}{i} and \element[][12]{CO}(2,0) and the water
amount for the 15 OH/IR stars in \citet{Schultheis2003} 
is also shown in Fig.~\ref{figure histoEW} (dotted line). Their
average equivalent widths are: 0.42 $\pm$ 1.56 \AA\ for EW(Ca), 2.50
$\pm$ 1.90 \AA\ for EW(Na) and 16.01 $\pm$ 5.16 \AA\ for EW(CO). The
equivalent widths for the OH/IR stars in this work are: 0.08 $\pm$
1.88 \AA\ for EW(Ca), 2.11 $\pm$ 1.27 \AA\ for EW(Na) and 15.60 $\pm$
4.69 \AA\ for EW(CO). The largest difference is seen for EW(Ca): a lot
of the OH/IR stars in this work and in \citet{Schultheis2003} have no
measurable \ion{Ca}{i} lines.\\


\section{Discussion}
\label{section discussion}

\subsection{Metallicities of OH/IR stars in the GC}
\label{section metallicities of OH/IR stars in the GC}
\citet{Schultheis2003} investigated the metallicity distribution of 107
ISOGAL sources in the GB. The sample consists
of different types of stars: non-variable giants, OH/IR stars, 
supergiants and LPVs. The metallicity
distribution is determined based on the equivalent line widths of
\ion{Na}{i}, \ion{Ca}{i} and \element[][12]{CO}(2,0)
\citep[see][]{Ramirez2000,Frogel2001,Schultheis2003}. The calibration
of the relation is based on giants in globular clusters in the range
$-1.8 < {\rm[Fe/H]} < -0.1$. The mean [Fe/H] value in
\citet{Schultheis2003} is consistent with previous chemical abundance
studies of the GB \citep{Schultheis2003}. Apparantly variable stars,
such as Miras and OH/IR stars, do not influence the peak in the metallicity
distribution determined in \citet{Schultheis2003}, although they might broaden
the distribution.\\
Applying the results of \citet{Schultheis2003} to our sample of OH/IR
stars could help to find a metallicity difference between the two different
groups of OH/IR stars in the GC as discussed in the introduction.\\
The equivalent line widths of \ion{Na}{i},
\ion{Ca}{i} and \element[][12]{CO}(2,0) are determined as
described in section \ref{section equivalent widths}. The values,
given in Tables \ref{table sample 28/06}, \ref{table sample 29/06} and
\ref{table sample 30/06}, are given in \AA. A positive
value indicates absorption and a negative value indicates
emission. Immediately one notices the rather large amount of negative
values for EW(Ca). One does not expect the \ion{Ca}{i} line and the
\ion{Na}{i} line to be in emission. Inspecting the
spectra for these stars (Fig. \ref{figure OH/IR 1} until
Fig. \ref{figure OH/IR 4}) shows that for a lot of stars the
\ion{Ca}{i} lines seem to disappear in the continuum. This effect is
also noticable for \ion{Na}{i} in some stars.\\

In the rest of this section we will discuss the physical effects that
can cause the \ion{Ca}{i} and sometimes the \ion{Na}{i} to appear
apparently in emission.

\subsection{Dust}
\label{section dust}

\begin{figure}
\begin{center}
\resizebox{\hsize}{!}{\includegraphics{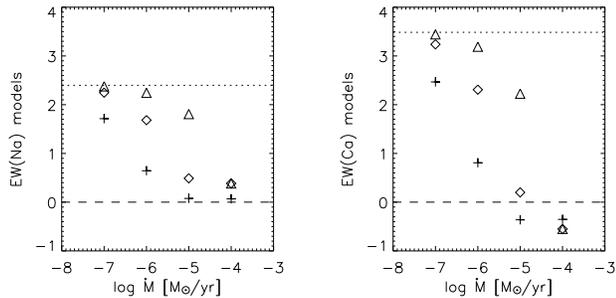}}
\caption{Equivalent line widths for the different models. The x-axis
  indicates the mass-loss rate, the y-axis the equivalent line
  widths for \ion{Na}{i} (left) and \ion{Ca}{i} (right). The different
  symbols indicate the different dust temperatures. Crosses: $T_{dust}
  = 1500$~K, diamonds: $T_{dust} = 1000$~K and triangles: $T_{dust} =
  750$~K. The dotted line indicates the equivalent line width for the
  reference line. The line disappears when the equivalent width is
  below the dashed line.}
\label{figure dust}
\end{center}
\end{figure}

OH/IR stars are AGB stars in their final phase on the AGB and are
believed to be progenitors of Planetary Nebulae
\citep[e.g.][]{Cohen2005,Habing1996}.
OH/IR stars have high mass-loss rates between 
$\dot{M} = 10^{-6}~\Msun/yr$ and $10^{-4}~\Msun/yr$. 
It is not unlikely that the dust formed around the star 
has an influence on the near-IR spectrum \citep{Tej2003}.\\
To study the effect of dust on the \ion{Na}{i} and \ion{Ca}{i} lines, the dust
radiative transfer model of \citet{GroenewegenPhDT} was used. In this
model, the radiative transfer equation and the radiative equilibrium
equation for the dust are solved simultaneously in spherical
geometry \citep{Groenewegen1994}. For the dust properties we assume
silicate dust \citep{Volk1988} of radius $a=0.02$~\micron\
and specific dust density of 
$\rho_{\rm d} = 2.0$~g/~cm$^3$. As 
input model we used a blackbody 
with $\Teff = 2500$~K and with the \ion{Ca}{i} and \ion{Na}{i} lines
imposed upon it. A dust-to-gas ratio of $\Psi = 0.01$ and an outflow
velocity of $v = 15$~km/s were assumed. For a grid in
mass-loss and dust temperature the models give an indication of the
dust-influence on these lines. Typical AGB mass-loss rates (between $\dot{M} =
10^{-7}~\Msun/yr$ and $10^{-4}~\Msun/yr$) and dust
temperatures ($T_{dust} = 1500, 1000$ and $750~K$) were used. The model
with no mass-loss is used as a reference model.\\

Fig. \ref{figure dust} shows the effect of the increasing dust amount
on the equivalent line widths of \ion{Na}{i} and \ion{Ca}{i}. For both
lines a similar trend can be seen in which the equivalent line widths
decrease for increasing mass loss rates. The effect is stronger 
for higher dust temperatures, except at the $10^{-4}~\Msun/yr$ mass loss
rate where the \ion{Na}{i} and \ion{Ca}{i} lines become undetectable for
the three dust temperatures. The effect is also stronger with increasing 
wavelength and impacts the \ion{Ca}{i} more than the \ion{Na}{i}.
Unfortunately, we do not know the 
temperature of the dust, neither do we know the mass-loss rate for
our individual sources. We can only demonstrate here the possible influence
of the dust on the measured near-infrared spectrum.
Surprisingly, negative values for EW(Ca) were measured for these
models. This is 
merely due to the chosen continuum: the change in the continuum's
slope is so drastic that it causes negative equivalent line width
measurements. \\

\subsection{Water content in OH/IR stars} 
\label{section water absorption in OH/IR stars}
\begin{figure}
\begin{center}
\resizebox{\hsize}{!}{\includegraphics{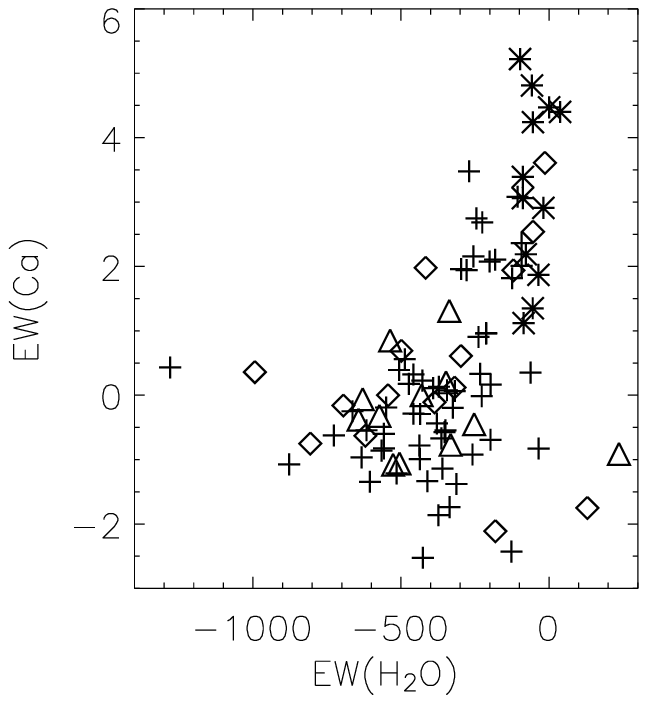}}
\resizebox{\hsize}{!}{\includegraphics{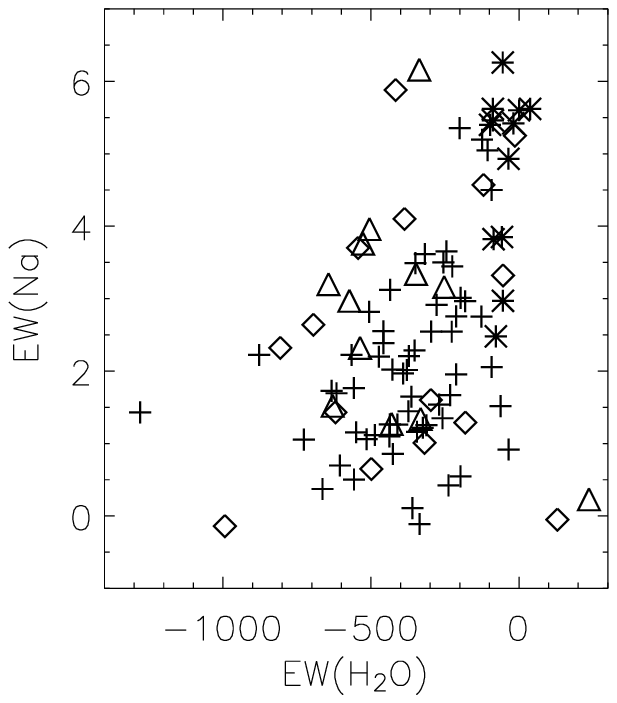}}
\caption{Upper panel: EW(Ca) vs EW(H$_2$O), lower panel: EW(Na) vs
EW(H$_2$O). The OH/IR in this work are the crosses, RGB
\citep{Schultheis2003}: stars, OH/IR \citep{Schultheis2003}: diamonds, LPV
\citep{Schultheis2003}: triangles. All equivalent line widths are
given in the same unit \AA.}
\label{figure water absorption effect}
\end{center}
\end{figure}

Water absorption lines can influence the near-IR medium resolution 
spectra of AGB stars severely in a way which depends
on the phase of the variable star as was shown in 
\citet{Tej2003}. The high resolution spectrum of $o$~Cet, a M
type Mira, in 
\citet{Wallace1996} shows that a lot of water lines are
situated at wavelengths where the \ion{Ca}{i} and \ion{Na}{i} lines 
are observed. In the case of the \ion{Ca}{i} line, the strong waterlines
are situated where the continuum of the line is determined. The water
lines depress the continuum in a medium resolution
spectrum and make the equivalent line width measurements unreliable. 
For a strong enough depression of the continuum, one can expect to 
start to see the \ion{Ca}{i} line  apparently in emission. 
In the case of the \ion{Na}{i} line, the water lines affect both the continuum
and the \ion{Na}{i} line, so that one can expect to see the EW(Na) decrease
because of a smaller contrast between the continuum and the line, similar
to what is seen due to the influence of the dust on the spectrum, as discussed
in the previous subsection. \\
These effects are reflected in Fig. \ref{figure water absorption effect}
where the  EW(Ca) and the EW(Na) can be seen in comparison with our
determination of the water absorption. We see that the RGB stars, with
little water absorption,  have higher EW(Ca) and the EW(Na) values
than the LPVs and OH/IR stars. The effect  seems stronger for
\ion{Ca}{i} where for EW(\HtwoO) below -300~\AA, we basically  have no
detections any more. For the
\ion{Na}{i} line the decrease is more gradual with increasing water
absorption. Of course the effect of dust and other factors
  (such as pulsation amplitude, \Teff) on the spectra is also
included in these figures and cannot be distinguished from the effect
that the water absorption has.\\

\subsection{Variability}
\label{section periodicity}

\begin{figure}
\begin{center}
\resizebox{\hsize}{!}{\includegraphics{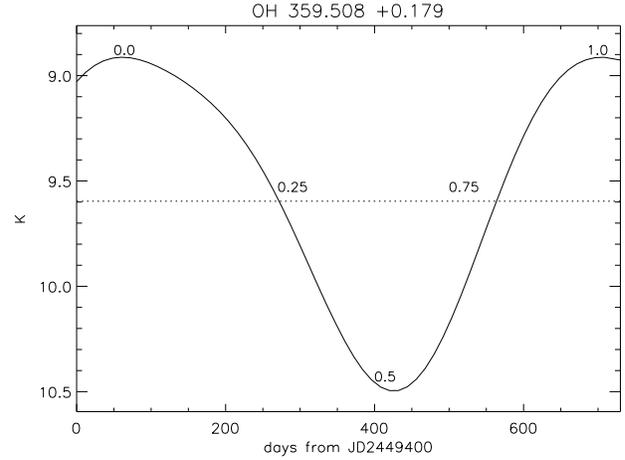}}
\caption{Determination of the phase $\phi$, $<K>$ is represented by
  the dotted horizontal line.}
\label{figure phase definition}
\end{center}
\end{figure}

\begin{figure}
\begin{center}
\resizebox{\hsize}{!}{\includegraphics{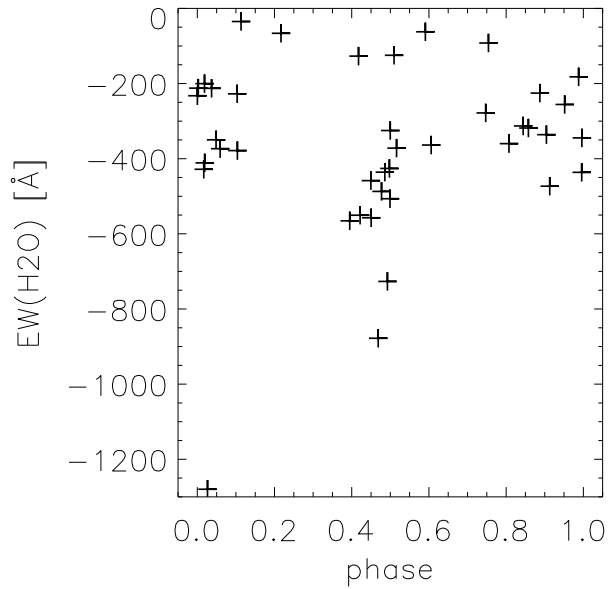}}
\caption{The phase at time of observation versus the equivalent line
  width of water absorption.}
\label{figure phase vs EW}
\end{center}
\end{figure}

Fig.~\ref{figure vergelijking OH/IR Mathias} shows clearly the
differences in the measured absorption for water for the stars we have
in common with \citet{Schultheis2003}. The water absorption is expected
to be largely correlated with the phase for a given star: the most
intense water features are to be seen at minimum light
\citep{Lancon2000,Bessel1996,Tej2003}.\\
\citet{Wood1998} determined periods for 80
GC OH/IR stars. We retrieved K light curves for 41 out of
the 50 stars 
in common with our sample. The lightcurves were extrapolated using the
fundamental period and the first harmonic \citep{Wood2004} to
determine the phase $(= \phi)$ when the star was observed in our
campaign. $\phi$ has been calculated in a non-traditional
  way. We calculated $\phi$ such that the maxiumum of the lightcurve
  is 0 and that the minimum is always at $\phi = 0.5$. When the
  lightcurve reaches $<K>$ the values of 0.25/0.75 were fixed. The
  other values for $\phi$ were then interpolated between the fixed
  ones (see Fig.~\ref{figure phase definition}).


 Fig.~\ref{figure phase vs EW} shows the
phase $\phi$ versus the equivalent line width of water. 
The expected correlation can be seen in this figure, but not as clear as
anticipated. Fig.~\ref{figure phase vs EW} indicates that the water
absorption is highest, except for one star, for $\phi=0.5$, which
indicates the light minimum. For the other phases there is a large
spread and no clear trend, which can be expected as the individual stars
have different parameters.

\subsection{Expansion velocities}
\label{section expansion velocities}

\begin{figure}
\begin{center}
\resizebox{\hsize}{!}{\includegraphics{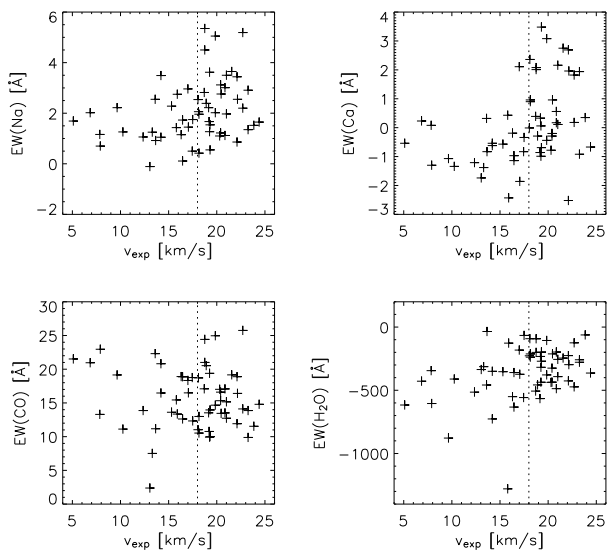}}
\caption{Comparison of the expansion velocities with the equivalent
  line widths of \ion{Na}{i}, \ion{Ca}{i}, \element[][12]{CO}(2,0) and
  the water absorption. The dotted vertical line separates the 2
  groups of OH/IR stars.}
\label{figure expansion velocities}
\end{center}
\end{figure}

\begin{table}
\caption{Mean values for the two groups of OH/IR stars.}
\label{table mean values 2 groups}
\begin{center}
\begin{tabular}{ccc}
\hline\hline
                        & $v_{\rm exp} < 18$ km/s & $v_{\rm exp} > 18$ km/s \\
\hline
\ion{Na}{i}             &  1.57 $\pm$ 0.70       &  2.44 $\pm$ 1.02\\
\ion{Ca}{i}             & -0.65 $\pm$ 0.69       &  0.67 $\pm$ 1.17\\
\element[][12]{CO}(2,0) & 15.73 $\pm$ 3.90       & 15.61 $\pm$ 3.44\\
H$_2$O & -452.90 $\pm$ 190.82 & -294.97 $\pm$ 110.31 \\
\hline
\end{tabular}
\end{center}
\end{table}

\citet{Lindqvist1992b} determined for their sample of OH/IR stars
close to the GC, the expansion velocities $v_{\rm exp}$. As mentioned
in section \ref{section introduction}, they divided the sample in two
groups based on the expansion velocities. The first group with the
lowest expansion velocities are kinematically different and show
larger dispersions in their radial velocities.\\

With our spectra we wanted to see whether we
could find metallicity differences between the two kinematically different
groups. As shown in the previous subsections, this is not possible
using the  relationship based on the equivalent line widths of
\ion{Na}{i}, \ion{Ca}{i} and \element[][12]{CO}(2,0) in
  nonvariable stars with low mass loss rates
\citep[see][]{Ramirez2000,Frogel2001,Schultheis2003}. We still wanted
to investigate whether we can find differences in the equivalent line
widths of the individual lines for the two groups.
Fig. \ref{figure expansion velocities} shows, for our sample of OH/IR
stars, the comparison between the expansion velocities and the
equivalent line widths of \ion{Na}{i}, \ion{Ca}{i},
\element[][12]{CO}(2,0) and the water absorption around
1.6~\micron. Table \ref{table mean values 2 groups} gives the mean
values with standard deviation for the measured equivalent line widths
and the water indication for the both groups. Fig. \ref{figure
  expansion velocities} and Table \ref{table mean values 2 groups}
seem to indicate that the OH/IR stars in the first group have smaller
EW(Na) and have a larger spread for the water 
absorption. Based on the numbers in Table \ref{table mean values 2
  groups} there is no difference
between the EW(CO) for both groups. Concerning \ion{Ca}{i}, the
majority of the OH/IR stars in the first group have no measurable
equivalent line widths. Since the water absorption has a larger spread
for this group and the dependence seen in Fig. \ref{figure water
  absorption effect} it is consistent that we measure for the majority
of these stars negative EW(Ca).\\
\citet{DecinPhDT} indicates that the linestrength of \HtwoO\ increases
with decreasing metallicity. This
could explain why we see stronger water absorption for the OH/IR
stars in the first group, provided these OH/IR stars are older and
have lower metallicicties than the stars in the second group, as
suggested by kinematics and outflow velocities.


\section{Conclusions}
\label{section conclusions}

We obtained near-IR data (1.53~--~2.52~\micron) with SOFI on the NTT
for 70 OH/IR stars located in the GC. The spectra were analysed based
on the equivalent line widths of \ion{Na}{i}, \ion{Ca}{i} and
\element[][12]{CO}(2,0). The curvature of the spectrum around
1.6~\micron\ gives us an indication of the water amount.
The  equivalent line widths of \ion{Na}{i}, \ion{Ca}{i} and
\element[][12]{CO}(2,0) were found to have low values in comparison to
the GC static giant stars. For a large fraction of the OH/IR stars, we
even found that \ion{Ca}{i} lines could not be detected.\\

We discuss different aspects which can influence the determination of the
equivalent line widths in the near-infrared spectra.
The OH/IR stars have a variable amount of water, which influences
especially the \ion{Ca}{i} lines. The water lines just besides the
\ion{Ca}{i} line depress the continuum, causing the \ion{Ca}{i} lines to
disappear in the spectrum. The effect is also noticeable for
\ion{Na}{i} but is less strong. \\
We also discuss the effect of the circumstellar dust on the
near-infrared spectrum.  Using the radiative transfer model
\citep{GroenewegenPhDT}, it became clear that for the highest
mass-loss rates the dust has the same effect on the \ion{Ca}{i} and
\ion{Na}{i} lines as the water content: the lines become weaker and in
the extreme case of mass loss rates in the order of $10^{-4}~\Msun/yr$
even disappear  in the continuum and are no longer measurable. For the
lower mass-loss rates ($\sim 10^{-6}~\Msun/yr$), the decrease of the
equivalent line widths depends strongly on the dust temperature. Since
we do not know the exact dust temperature and the mass-loss rates of
these OH/IR stars, we cannot distinguish between the possible effects
which weaken the lines.\\
The different effects discussed above prevent us from finding a clear
distinction between the two groups of OH/IR stars.

\begin{acknowledgements}
M.S. is supported by an APART fellowship. We want to thank P.R.~Wood,
who kindly provided us with the K~lightcurves for the OH/IR stars.
\end{acknowledgements}


\appendix
\section{Tables}
\begin{sidewaystable*}[h]
\begin{minipage}[t][180mm]{\textwidth}
\caption{\label{table sample 28/06} Log of the stars observed on 28/06/03.}
\centering
\begin{tabular}{llllrrrrrr}
\hline \hline
Name & RA & Dec & References \footnote{ Li: \citet{Lindqvist1992b},
  Sj: \citet{Sjouwerman1998} and \citet{Ortiz2002}, B98:
  \citet{Blommaert1998}, A: \citet{Schultheis2003}, HV:
  \citet{vanLangevelde1992}, R97: \citet{Ramirez1997}, R00:
  \citet{Ramirez2000}, 
  L00: \citet{Lancon2000}, and S98: \citet{Sjouwerman1998}.}
& EW(Na) & EW(Ca) & EW(CO) & \HtwoO & P \footnote{Periods are taken
  from \citet{Wood1998}.}
& $v_{\rm exp}$\\ 
 & & & & [\AA] & [\AA] & [\AA] & [\AA] & [days] & [m/s]\\
\hline
\object{OH 359.576 +0.091}  &  17 44 57.80  &  -29 20 42.5  &            Li012, A29  &   2.50  &   0.47  &  15.13  &          &  672.0  &  18.6  \\
\object{OH 359.598 +0.000}  &  17 44 39.71  &  -29 16 46.1  &            Li014, A10  &   1.67  &   0.33  &  10.77  & -232.39  &  664.0  &  19.2  \\
\object{OH 359.669 -0.019}  &  17 44 54.16  &  -29 13 44.9  &            Li020, A23  &   0.50  &  -0.83  &  18.76  & -557.33  &  481.0  &  17.5  \\
\object{OH 359.675 +0.069}  &  17 44 34.33  &  -29 10 38.6  &  Li021, A07, B98, S98  &   5.35  &   2.08  &  24.42  & -200.35  &  698.0  &  18.8  \\
\object{OH 359.678 -0.024}  &  17 44 56.83  &  -29 13 25.5  &            Li022, B98  &   3.65  &   2.75  &  19.16  & -245.18  &         &  21.5  \\
\object{OH 359.681 -0.095}  &  17 45 16.43  &  -29 15 37.6  &       Li023, A51, S98  &   2.76  &   2.78  &  17.01  &  -43.64  &  759.0  &  19.3  \\
\object{OH 359.711 -0.100}  &  17 45 19.26  &  -29 14 01.1  &       Li025, A58, S98  &   3.62  &   0.06  &  19.39  & -318.34  &  686.0  &  19.2  \\
\object{OH 359.748 +0.274}  &  17 43 57.05  &  -29 00 28.4  &                 Li030  &  -0.11  &  -1.74  &   2.38  & -335.99  &  437.0  &  13.1  \\
\object{OH 359.755 +0.061}  &  17 44 47.94  &  -29 06 49.9  &  Li031, A14, B98, S98  &   4.50  &   2.01  &  21.01  &  -92.73  &    0.0  &  18.8  \\
\object{OH 359.757 -0.136}  &  17 45 34.41  &  -29 12 54.1  &       Sj002, A71, S98  &   1.10  &  -0.78  &  17.02  & -437.67  &         &  20.3  \\
\object{OH 359.762 +0.120}  &  17 44 34.96  &  -29 04 35.7  &  Li033, B98, E51, S98  &   2.28  &  -0.57  &  13.62  & -352.97  &         &  15.3  \\
\object{OH 359.765 +0.082}  &  17 44 44.43  &  -29 05 37.9  &       Li035, A12, S98  &   2.75  &  -2.43  &  13.35  & -126.89  &  552.0  &  15.9  \\
\object{OH 359.768 -0.207}  &  17 45 52.62  &  -29 14 30.4  &                 Li036  &   1.27  &  -0.99  &   9.92  & -436.23  &  602.0  &  19.2  \\
\object{OH 359.783 -0.392}  &  17 46 38.12  &  -29 19 30.7  &                 Li039  &   1.25  &  -1.38  &   7.51  & -312.88  &  559.0  &  13.3  \\
\object{OH 359.797 -0.025}  &  17 45 14.27  &  -29 07 20.8  &       Sj004, A52, S98  &   1.54  &   3.48  &  13.96  & -269.95  &         &  19.3  \\
\object{OH 359.799 -0.090}  &  17 45 29.50  &  -29 09 16.0  &       Li040, B98, S98  &   1.11  &  -0.51  &  15.97  &          &         &  18.1  \\
\object{OH 359.800 +0.165}  &  17 44 30.11  &  -29 01 14.3  &                 Li041  &   3.49  &  -0.54  &  16.47  & -349.81  &  461.0  &  14.2  \\
\object{OH 359.803 -0.248}  &  17 46 07.10  &  -29 13 56.1  &            Li043, S98  &   4.72  &   3.94  &  20.13  &  -25.58  &  803.0  &  18.5  \\
\object{OH 359.810 -0.070}  &  17 45 26.35  &  -29 08 03.9  &  Li044, A64, B98, S98  &   2.55  &   1.96  &  16.41  & -297.19  &         &  22.1  \\
\object{OH 359.825 -0.024}  &  17 45 17.83  &  -29 05 53.3  &       Li047, B98, S98  &   0.37  &  -0.25  &  19.95  & -663.81  &         &        \\
\object{OH 359.837 +0.030}  &  17 45 06.98  &  -29 03 34.9  &  Li048, A43, B98, S98  &   1.16  &   0.08  &  13.30  & -344.67  &  402.0  &   7.8  \\
\object{OH 359.838 +0.053}  &  17 45 01.70  &  -29 02 49.9  &            Sj011, A36  &   1.06  &  -1.21  &  13.87  & -514.62  &         &  12.4  \\
\object{OH 359.889 +0.361}  &  17 43 56.94  &  -28 50 30.8  &                 Li054  &   2.54  &  -0.01  &  11.03  & -227.34  &  389.0  &  18.1  \\
\object{OH 359.906 -0.041}  &  17 45 33.17  &  -29 02 18.4  &              B98, S98  &   5.05  &   3.08  &  24.96  & -106.08  &         &  19.8  \\
\object{OH 359.943 +0.260}  &  17 44 28.17  &  -28 50 55.8  &       Li064, B98, S98  &   1.75  &  -0.34  &  12.33  &  -66.26  &  692.0  &  17.5  \\
  \object{OH 0.001 +0.352}  &  17 44 14.95  &  -28 45 06.0  &            Li076, S98  &   2.22  &  -1.07  &  19.15  & -877.91  &  477.0  &   9.7  \\
  \object{OH 0.019 +0.345}  &  17 44 19.23  &  -28 44 21.8  &                 Li079  &   1.52  &   0.35  &  11.54  &  -62.35  &  701.0  &  23.9  \\
  \object{OH 0.138 -0.136}  &  17 46 28.71  &  -28 53 19.5  &            Li094, S98  &   3.12  &  -0.29  &  16.60  & -435.76  &  622.0  &  20.4  \\
  \object{OH 0.173 +0.211}  &  17 45 12.45  &  -28 40 44.4  &            Li096, S98  &   1.45  &  -1.86  &  16.49  & -373.41  &  514.0  &  17.0  \\
  \object{OH 0.200 +0.233}  &  17 45 11.44  &  -28 38 43.2  &                 Li101  &   1.43  &   0.43  &  15.44  &-1279.67  &  825.0  &  15.8  \\
  \object{OH 0.221 +0.168}  &  17 45 29.02  &  -28 39 38.5  &                 Li104  &   2.22  &  -0.86  &  13.53  & -565.55  &  697.0  &  19.2  \\
        \object{HD 130223}  &  14 47 53.62  &  -32 14 49.5  &                   R97  &   2.72  &   2.48  &  17.25  &  -11.74  &         &        \\
        \object{HD 133336}  &  15 04 52.37  &  -31 16 44.7  &                   R97  &   2.32  &   2.37  &  15.65  &  -11.83  &         &        \\
        \object{HD 181012}  &  19 20 13.43  &  -36 13 39.8  &                   R97  &   2.10  &   2.21  &  17.63  &  -17.89  &         &        \\
        \object{HD 224715}  &  00 00 09.82  &  -35 57 36.8  &                   R97  &   1.84  &   2.04  &  14.44  &  -33.38  &         &        \\
           \object{WX Psc}  &  01 06 25.98  &  +12 35 53.0  &                   L00  &   2.10  &   0.65  &  15.06  & -290.88  &         &        \\
\hline
\end{tabular}
\vfill
\end{minipage}
\end{sidewaystable*}

\begin{sidewaystable*}[h]
\begin{minipage}[t][180mm]{\textwidth}
\caption{\label{table sample 29/06} Log of the stars observed on 29/06/03.}
\centering
\begin{tabular}{llllrrrrrr}
\hline \hline
Name & RA & Dec & References \footnote{See Table \ref{table sample 28/06}.}
& EW(Na) & EW(Ca) & EW(CO) & \HtwoO & P \footnote{See Table \ref{table sample 28/06}.}
& $v_{\rm exp}$\\ 
 & & & & [\AA] & [\AA] & [\AA] & [\AA] & [days] & [m/s]\\
\hline
\object{OH 359.437 -0.051}  &  17 44 28.30  &  -29 26 35.0  &                 Li004  &   4.77  &   3.05  &  19.80  &  -45.15  &    0.0  &  15.3  \\
\object{OH 359.508 +0.179}  &  17 43 44.74  &  -29 15 44.6  &                 Li009  &   2.02  &  -0.44  &  14.67  & -378.36  &  644.0  &  19.8  \\
\object{OH 359.513 +0.174}  &  17 43 46.73  &  -29 15 38.4  &                 Li010  &   1.26  &  -1.34  &  11.12  & -411.05  &  461.0  &  10.2  \\
\object{OH 359.634 -0.195}  &  17 45 30.45  &  -29 21 00.0  &                 Li016  &   2.21  &   0.13  &  15.22  & -371.80  &  501.0  &   0.0  \\
\object{OH 359.636 -0.108}  &  17 45 10.39  &  -29 18 12.3  &                 Li017  &   0.86  &  -2.52  &  11.92  & -425.97  &  847.0  &  22.1  \\
\object{OH 359.640 -0.084}  &  17 45 05.29  &  -29 17 13.4  &                 Li018  &   2.55  &   0.32  &  22.31  & -458.27  &  546.0  &  13.6  \\
\object{OH 359.684 -0.104}  &  17 45 16.43  &  -29 15 37.6  &            Li024, S98  &   1.15  &  -0.19  &  18.91  & -550.36  &  535.0  &  16.3  \\
\object{OH 359.716 -0.070}  &  17 45 13.00  &  -29 12 54.0  &            Li026, S98  &   5.19  &   1.82  &  25.75  & -124.72  &  691.0  &  22.7  \\
\object{OH 359.719 +0.025}  &  17 44 51.25  &  -29 09 45.3  &                 Li027  &   2.20  &   0.18  &  14.11  & -473.12  &  669.0  &  22.7  \\
\object{OH 359.760 +0.072}  &  17 44 45.96  &  -29 06 13.6  &                 Li032  &   1.12  &   0.56  &  17.08  & -487.08  &  676.0  &  20.9  \\
\object{OH 359.763 -0.042}  &  17 45 13.07  &  -29 09 36.3  &            Li034, S98  &   1.05  &  -0.62  &  20.81  & -726.67  &  453.0  &  14.2  \\
  \object{OH 0.018 +0.156}  &  17 45 03.30  &  -28 50 22.0  &            Li078, S98  &   2.02  &   0.23  &  20.96  & -428.16  &  423.0  &   6.9  \\
  \object{OH 0.036 -0.182}  &  17 46 24.87  &  -29 00 01.4  &            Li080, S98  &   1.98  &  -1.61  &   7.64  &-4085.92  &  660.0  &  18.8  \\
  \object{OH 0.040 -0.056}  &  17 45 56.08  &  -28 55 52.0  &              B98, S98  &   3.01  &   0.17  &  13.52  & -197.38  &         &  20.8  \\
  \object{OH 0.060 -0.018}  &  17 45 50.07  &  -28 53 38.1  &                   B98  &   0.55  &  -0.69  &  10.01  & -197.71  &         &  19.3  \\
  \object{OH 0.076 +0.146}  &  17 45 13.91  &  -28 47 43.2  &       Li086, B98, S98  &   1.97  &   0.11  &  12.73  & -391.78  &         &  21.0  \\
  \object{OH 0.129 +0.103}  &  17 45 31.45  &  -28 46 22.1  &       Li091, B98, S98  &   1.77  &  -0.60  &  21.15  & -557.53  &         &        \\
  \object{OH 0.142 +0.026}  &  17 45 51.33  &  -28 48 06.8  &                   B98  &   1.35  &  -0.92  &   9.88  & -258.59  &         &  23.2  \\
  \object{OH 0.178 -0.055}  &  17 46 13.85  &  -28 48 50.4  &                   B98  &   1.73  &  -0.97  &  18.87  & -632.77  &         &  16.5  \\
  \object{OH 0.225 -0.055}  &  17 46 22.11  &  -28 46 22.6  &            Li105, B98  &   0.11  &  -1.14  &  12.62  & -359.80  &  521.0  &  16.5  \\
  \object{OH 0.241 -0.014}  &  17 46 14.96  &  -28 44 17.3  &                 Li106  &   2.82  &   0.39  &  17.09  & -506.51  &  535.0  &  18.7  \\
  \object{OH 0.265 -0.078}  &  17 46 33.12  &  -28 45 00.7  &                 Li108  &   2.06  &   2.36  &  18.71  &  -92.13  &  595.0  &  18.1  \\
  \object{OH 0.274 +0.086}  &  17 45 56.08  &  -28 39 27.5  &                 Li109  &   1.65  &  -0.67  &  14.80  & -363.65  &  706.0  &  24.4  \\
  \object{OH 0.307 -0.176}  &  17 47 02.17  &  -28 45 55.8  &                 Li110  &   1.22  &  -0.20  &  15.33  & -324.90  &  657.0  &  20.4  \\
  \object{OH 0.336 -0.027}  &  17 46 31.30  &  -28 39 48.0  &                 Li113  &   2.96  &   2.11  &  18.33  & -182.05  &  514.0  &  17.0  \\
        \object{HD 109467}  &  12 35 04.30  &  -28 46 39.6  &                   R97  &   2.73  &   1.84  &  17.68  &  -15.21  &         &        \\
        \object{HD 115141}  &  13 15 49.54  &  -40 03 17.1  &                   R97  &   0.93  &   0.29  &   7.28  &   -8.41  &         &        \\
        \object{HD 117927}  &  13 34 15.34  &  -34 23 15.3  &                   R97  &   0.90  &   0.19  &   6.03  &   -2.47  &         &        \\
        \object{HD 121870}  &  13 58 53.01  &  -32 47 32.5  &                   R97  &   2.87  &   1.69  &  17.03  &  -24.24  &         &        \\
        \object{AFGL 1686}  &  14 11 17.61  &  -07 44 49.9  &                   L00  &   1.96  &  -0.79  &  18.87  & -528.16  &         &        \\
            \object{R Phe}  &  23 56 27.57  &  -49 47 12.5  &                   L00  &   1.86  &   0.23  &  12.89  & -118.80  &         &        \\
            \object{S Phe}  &  23 59 04.57  &  -56 34 32.3  &                   L00  &   2.17  &   1.48  &  17.56  &  -15.30  &         &        \\
           \object{SV Tel}  &  18 56 17.94  &  -49 29 09.7  &                   L00  &  -1.20  &   0.29  &  11.69  & -481.43  &         &        \\
            \object{U Equ}  &  20 57 16.28  &  +02 58 44.6  &                   L00  &  -1.65  &   0.70  &  -2.35  & -109.43  &         &        \\
            \object{Z Aql}  &  20 15 11.04  &  -06 09 04.0  &                   L00  &   0.12  &   0.53  &  12.65  & -201.51  &         &        \\
\hline
\end{tabular}
\vfill
\end{minipage}
\end{sidewaystable*}

\begin{sidewaystable*}[h]
\begin{minipage}[t][180mm]{\textwidth}
\caption{\label{table sample 30/06} Log of the stars observed on 30/06/03.}
\centering
\begin{tabular}{llllrrrrrr}
\hline \hline
Name & RA & Dec & References \footnote{See Table \ref{table sample 28/06}.}
& EW(Na) & EW(Ca) & EW(CO) & \HtwoO & P \footnote{See Table \ref{table sample 28/06}.}
& $v_{\rm exp}$\\ 
 & & & & [\AA] & [\AA] & [\AA] & [\AA] & [days] & [m/s]\\
\hline
\object{OH 359.746 +0.134}  &  17 44 29.42  &  -29 04 58.6  &            Li029, S98  &   0.70  &  -1.30  &  22.96  & -604.71  &         &   7.9  \\
\object{OH 359.778 +0.010}  &  17 45 03.20  &  -29 07 12.6  &            Li038, S98  &   3.50  &   2.16  &  15.16  & -255.61  &  557.0  &  21.0  \\
\object{OH 359.814 -0.162}  &  17 45 48.48  &  -29 10 45.2  &            Li045, S98  &   2.76  &   0.96  &  13.45  & -212.24  &  554.0  &  20.5  \\
\object{OH 359.836 +0.119}  &  17 44 45.97  &  -29 00 50.9  &             Sj009, HV  &  -1.31  &   1.01  &  14.06  & -126.84  &         &        \\
\object{OH 359.855 -0.078}  &  17 45 34.79  &  -29 06 02.7  &                 Li050  &   3.44  &   2.69  &  18.89  & -225.50  &  617.0  &  22.1  \\
\object{OH 359.864 +0.056}  &  17 45 04.71  &  -29 01 24.9  &        Sj013, A38, HV  &   0.09  &   0.57  &  22.89  & -660.83  &         &        \\
\object{OH 359.918 -0.055}  &  17 45 38.46  &  -29 02 03.9  &              B98, S98  &   0.42  &   0.90  &  10.53  & -238.09  &         &  18.1  \\
\object{OH 359.943 -0.055}  &  17 44 28.20  &  -28 50 55.0  &                   S98  &   2.49  &   1.66  &  14.73  &   31.60  &         &        \\
  \object{OH 0.017 +0.156}  &  17 45 03.65  &  -28 50 27.7  &                   S98  &   2.39  &  -0.29  &  20.53  & -457.78  &         &  18.9  \\
  \object{OH 0.037 -0.003}  &  17 45 43.16  &  -28 54 21.8  &                   S98  &   1.69  &  -0.54  &  21.52  & -616.51  &         &   5.1  \\
  \object{OH 0.335 -0.180}  &  17 47 07.34  &  -28 44 28.3  &               HV, S98  &   3.33  &   1.65  &  17.06  &   -2.82  &         &        \\
  \object{OH 0.352 +0.175}  &  17 45 46.57  &  -28 32 39.5  &                 Li115  &   1.96  &   0.96  &  12.99  & -212.56  &  661.0  &  18.1  \\
  \object{OH 0.379 +0.159}  &  17 45 54.18  &  -28 31 46.0  &                 Li116  &   3.48  &  -0.37  &   2.75  &          &  985.0  &  15.3  \\
  \object{OH 0.395 +0.008}  &  17 46 31.48  &  -28 35 37.3  &                 Li117  &   0.92  &  -0.83  &  11.17  &  -34.99  &  461.0  &  13.7  \\
  \object{OH 0.447 -0.006}  &  17 46 42.29  &  -28 33 26.1  &                 Li120  &  -0.33  &  -9.57  &  14.46  &          &  445.0  &  13.1  \\
  \object{OH 0.452 +0.046}  &  17 46 30.70  &  -28 31 31.0  &                 Li121  &   1.96  &   1.47  &  14.14  &  -36.33  &  339.0  &  10.2  \\
  \object{OH 0.536 -0.130}  &  17 47 24.02  &  -28 32 42.9  &                 Li127  &   2.91  &   1.94  &  13.87  & -278.26  &  669.0  &  23.2  \\
                            &  17 44 23.80  &  -29 08 55.3  &                   A03  &   5.07  &   2.53  &  18.98  &  -10.57  &         &        \\
                            &  17 44 48.60  &  -29 00 13.2  &                   A16  &   6.46  &   1.16  &  20.84  &   -5.45  &         &        \\
                            &  17 44 53.10  &  -28 59 46.5  &                   A22  &   5.28  &   4.07  &  18.13  &   -6.80  &         &        \\
                            &  17 45 09.80  &  -29 05 17.8  &                   A45  &   3.28  &   0.68  &  16.53  &  -55.01  &         &        \\
                            &  17 45 27.50  &  -29 04 39.9  &                   A66  &   5.39  &   4.18  &  25.26  &  -58.05  &         &        \\
           \object{BMB 55}  &  18 03 08.16  &  -29 57 47.4  &                   R00  &   3.42  &   2.31  &  19.45  &  -90.37  &         &        \\
            \object{B-138}  &  18 03 46.00  &  -29 59 12.1  &                   R00  &   4.40  &   2.61  &  21.97  &  -48.03  &         &        \\
          \object{BMB 205}  &  18 03 54.85  &  -30 04 19.2  &                   R00  &   4.04  &   3.05  &  21.95  &  -57.11  &         &        \\
           \object{BMB 78}  &  18 03 15.54  &  -29 51 09.0  &                   R00  &   2.62  &   2.36  &  18.30  &  -28.21  &         &        \\
          \object{BMB 289}  &  18 04 22.72  &  -29 54 50.6  &                   R00  &   3.54  &   2.78  &  22.39  & -111.15  &         &        \\
            \object{B-124}  &  18 04 43.73  &  -30 05 15.3  &
	    R00  &   3.25  &   2.60  &  19.47  &  -33.46  &         &
	          \\
\hline
\end{tabular}
\vfill
\end{minipage}
\end{sidewaystable*}

\begin{table}[h]
  \caption{Contribution of other species to the equivalent line width
  measurement of \ion{Na}{i}. Based on \citet{Wallace1996}.}
  \label{table linelist Na}
  \begin{center}
    \begin{tabular}{ccc}\hline\hline
      \ion{Na}{i} feature & \ion{Na}{i} continuum \# 1 & \ion{Na}{i}
      continuum \# 2\\ \hline
      \element{CN}(0,2)  & \element{CN}(0,2) & \element{CN}(0,2)  \\
      \element{CN}(1,3)  & \element{CN}(1,3) & \element{CN}(1,3)  \\
      \element{CN}(2,4)  & \element{CN}(2,4) & \element{CN}(2,4)  \\
      \element{CN} blend & \element{Si}      & \element{CN} blend \\
      \element{V}        & \element{Fe}      &                    \\
      \element{Na}       &                   &                    \\
      \element{Fe}       &                   &                    \\
      \element{Si}       &                   &                    \\
      \element{Sc}       &                   &                    \\
      \hline
    \end{tabular}
  \end{center}
\end{table}

\begin{table}[h]
  \caption{Contribution of other species to the equivalent line width
  measurement of \ion{Ca}{i}. Based on \citet{Wallace1996}.}
  \label{table linelist Ca}
  \begin{center}
    \begin{tabular}{ccc}\hline\hline
      \ion{Ca}{i} feature & \ion{Ca}{i} continuum \# 1 & \ion{Ca}{i}
      continuum \# 2\\ \hline
      \element{CN}(0,2)  & \element{CN}(0,2)  & \element{CN}(2,4) \\
      \element{CN}(1,3)  & \element{CN}(1,3)  & \element{HF}      \\
      \element{CN}(2,4)  & \element{CN}(2,4)  & \element{S}       \\
      \element{CN} blend & \element{CN} blend &                   \\
      \element{Si}       & \element{Si}       &                   \\
      \element{Sc}       & \element{Fe}       &                   \\
      \element{Ca}       & \element{V}        &                   \\
      \element{Ti}       &                    &                   \\
      \element{Fe}       &                    &                   \\
      \element{Ni}       &                    &                   \\
      \hline
    \end{tabular}
  \end{center}
\end{table}

\begin{table}[h]
  \caption{Contribution of other species to the equivalent line width
  measurement of \element[][12]{CO}(2,0). Based on \citet{Wallace1996}.}
  \label{table linelist CO}
  \begin{center}
    \begin{tabular}{ccc}\hline\hline
      \element[][12]{CO}(2,0) & \element[][12]{CO}(2,0) &
      \element[][12]{CO}(2,0) \\ 
      feature & continuum \#1 & continuum \# 2 \\ \hline
      \element{CN}(0,2) & \element{CN}(0,2) & \element{CN}(0,2)  \\
      \element{CN}(1,3) & \element{CN}(2,4) & \element{CN}(1,3)  \\
      \element{CN}(2,4) & \element{Si}      & \element{CN}(2,4)  \\
      \element{Ti}      &                   & \element{CN} blend \\
      \element{HF}      &                   & \element{Ti}       \\
      \element{CO}(2,0) &                   & \element{HF}       \\
      \hline
    \end{tabular}
  \end{center}
\end{table}

\section{Near-IR Spectra}
\begin{center}
\begin{figure*}[h]
\resizebox{\hsize}{!}{\includegraphics{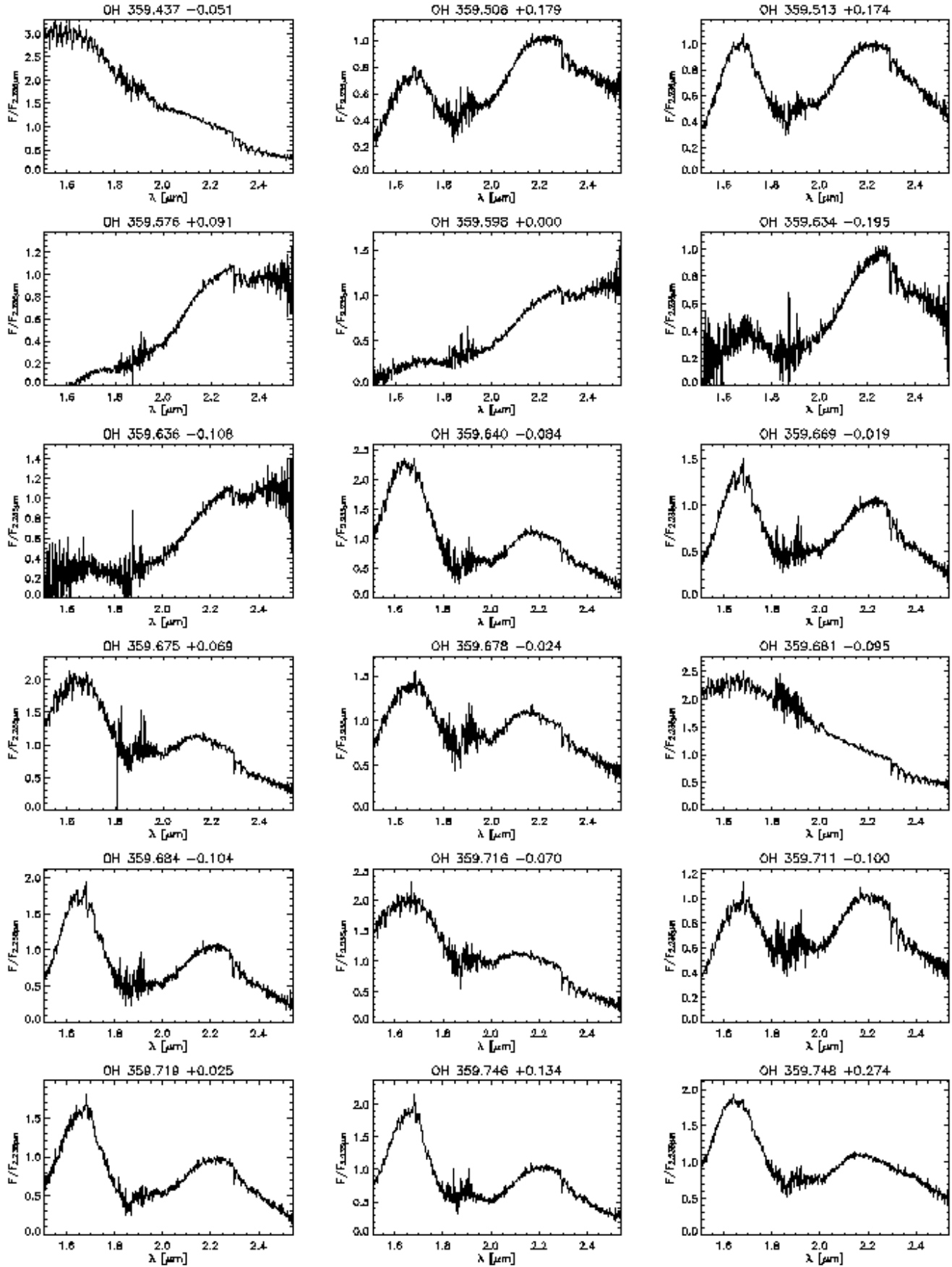}}
\caption{OH/IR stars. \object{OH 359.437-0.051} is possibly a mismatch.}
\label{figure OH/IR 1}
\end{figure*}
\begin{figure*}[h]
\resizebox{\hsize}{!}{\includegraphics{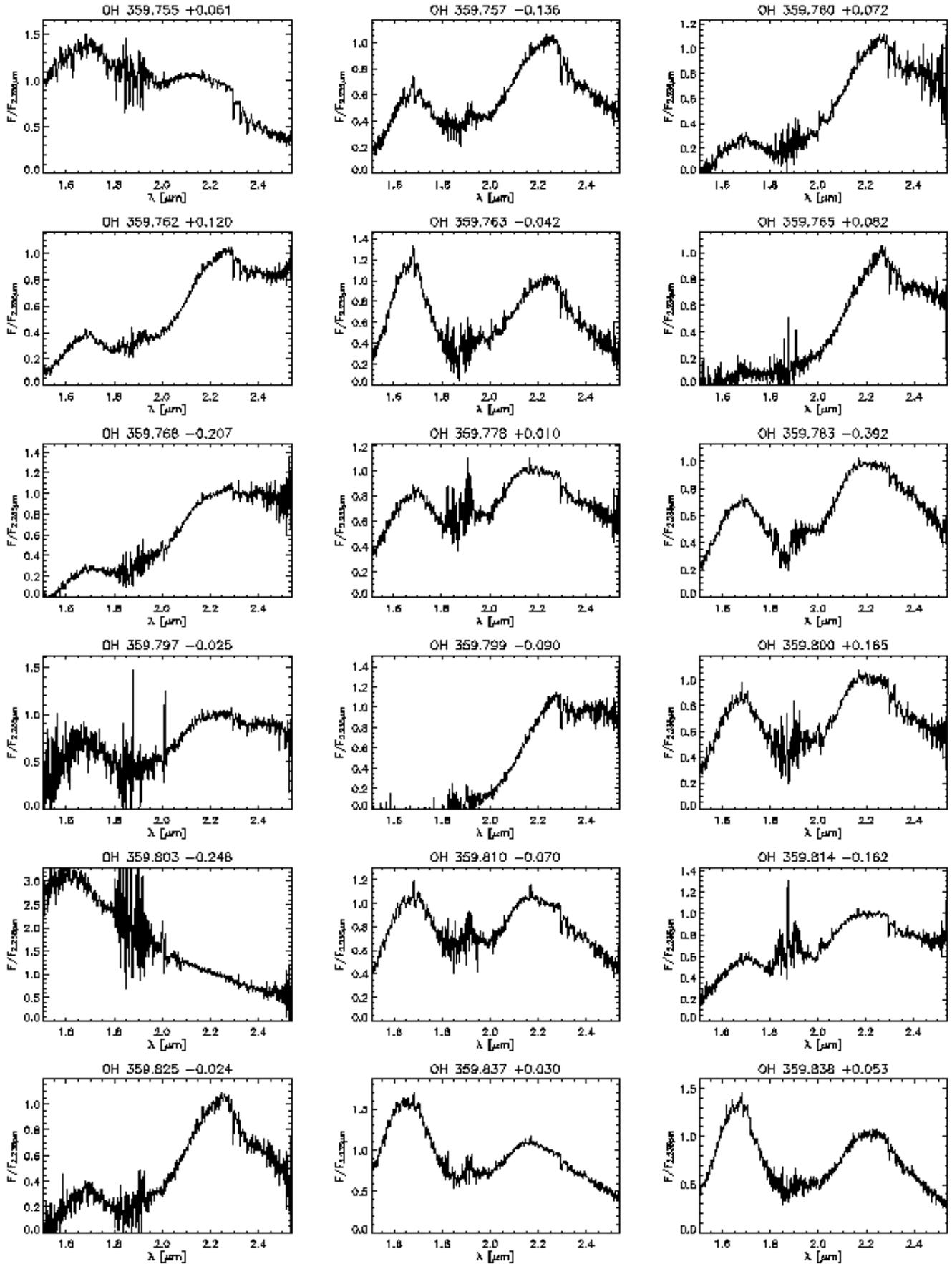}}
\caption{OH/IR stars - continue. \object{OH 359.803-0.248} is possibly
  a mismatch.}
\label{figure OH/IR 2}
\end{figure*}
\begin{figure*}[h]
\resizebox{\hsize}{!}{\includegraphics{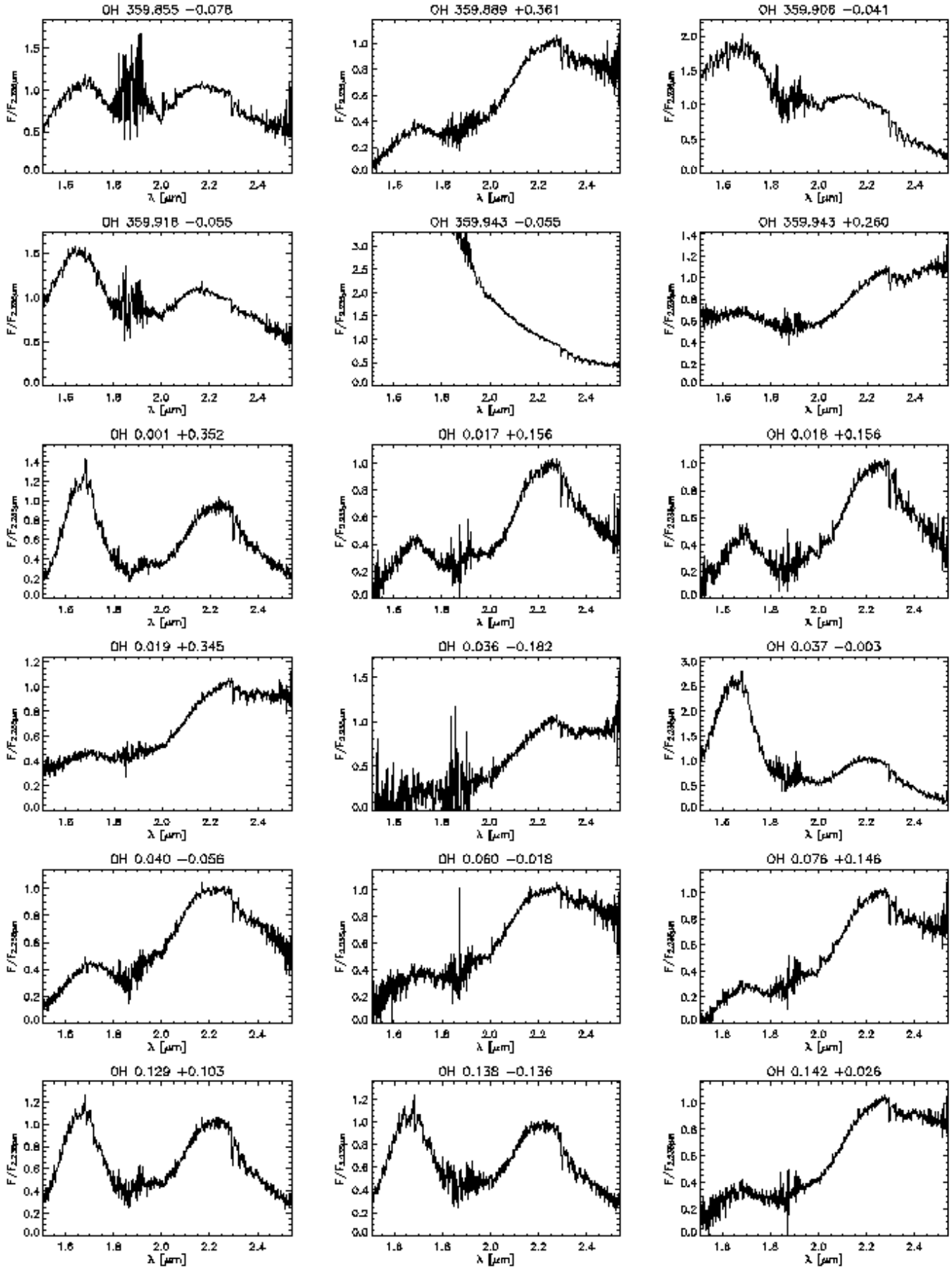}}
\caption{OH/IR stars - continue. \object{OH 359.943-0.055} is
  possibly a mismatch.}
\label{figure OH/IR 3}
\end{figure*}
\begin{figure*}[h]
\resizebox{\hsize}{!}{\includegraphics{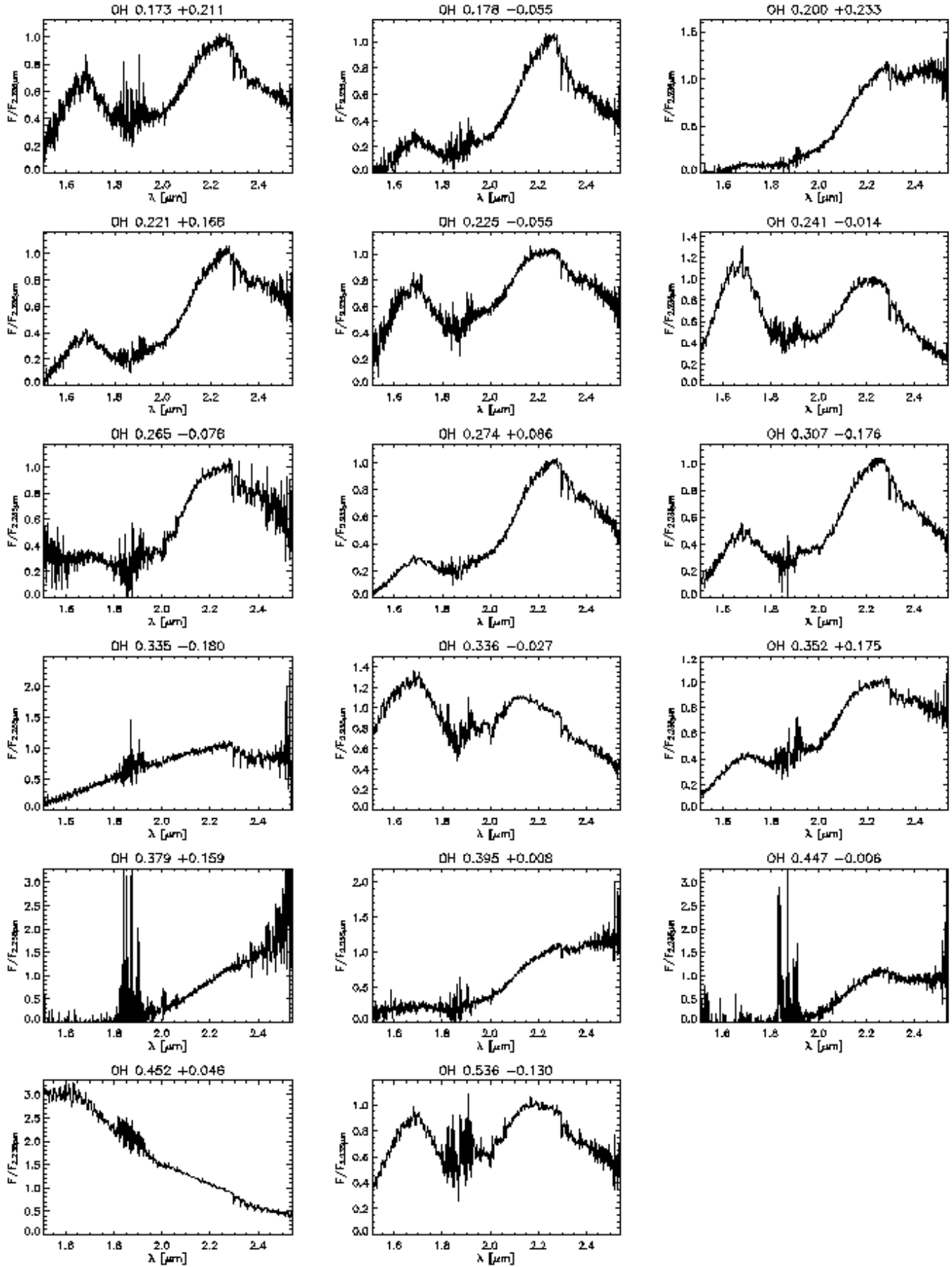}}
\caption{OH/IR stars - continue. \object{OH 0.452+0.046} is possibly a
  mismatch.}
\label{figure OH/IR 4}
\end{figure*}

\begin{figure*}[h]
\resizebox{\hsize}{!}{\includegraphics{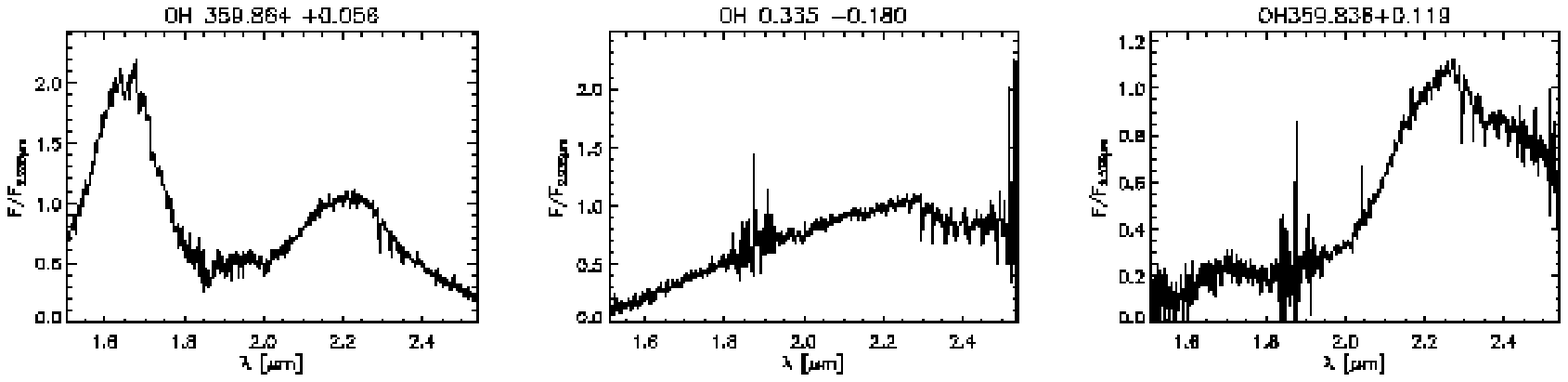}}
\caption{High velocity OH/IR stars \citep{vanLangevelde1992}.}
\label{figure HV}
\end{figure*}

\begin{figure*}
\resizebox{\hsize}{!}{\includegraphics{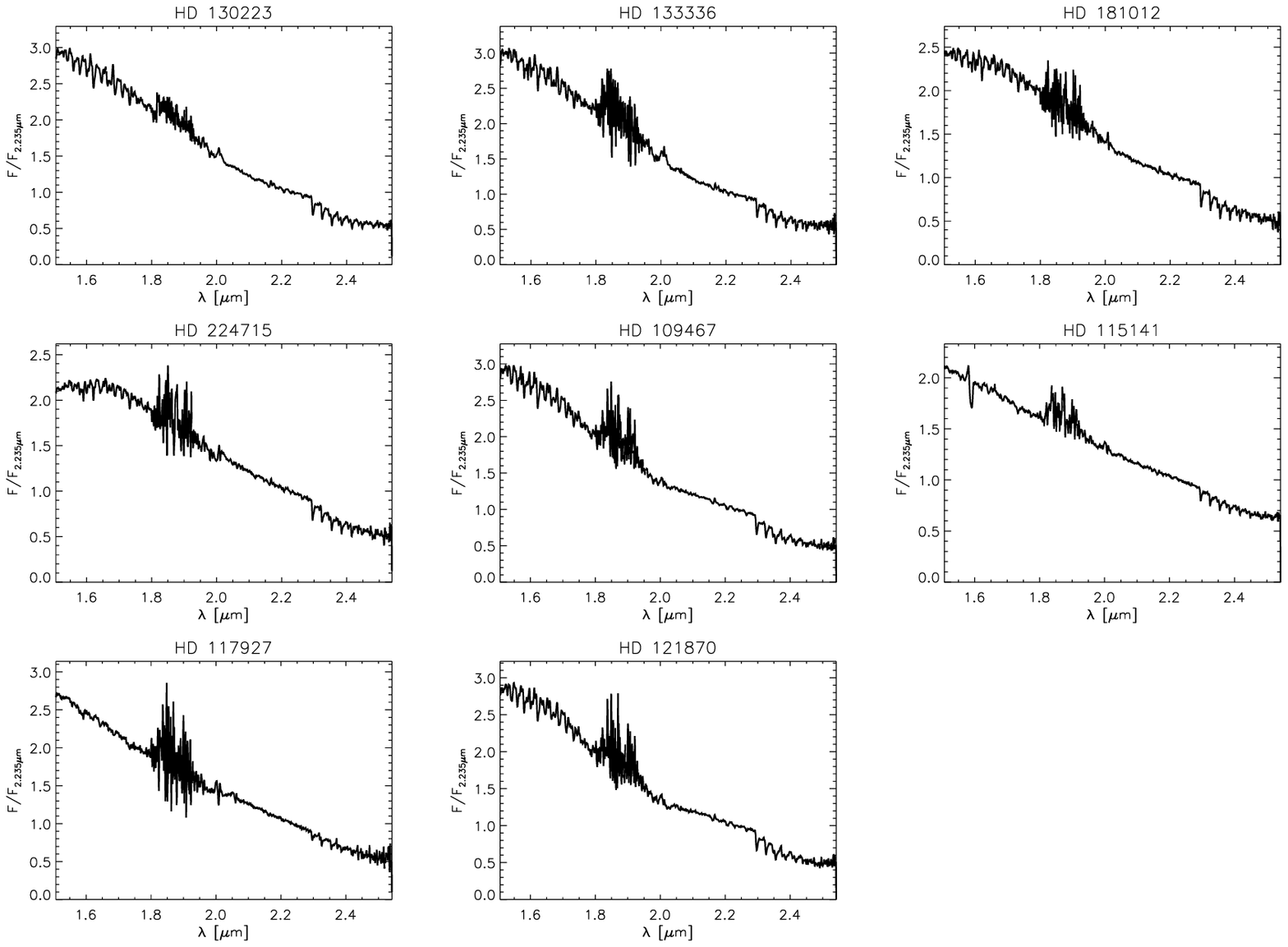}}
\caption{Stars in common with \citet{Ramirez1997}.}
\label{figure Ramirez1997}
\end{figure*}

\begin{figure*}[h]
\resizebox{\hsize}{!}{\includegraphics{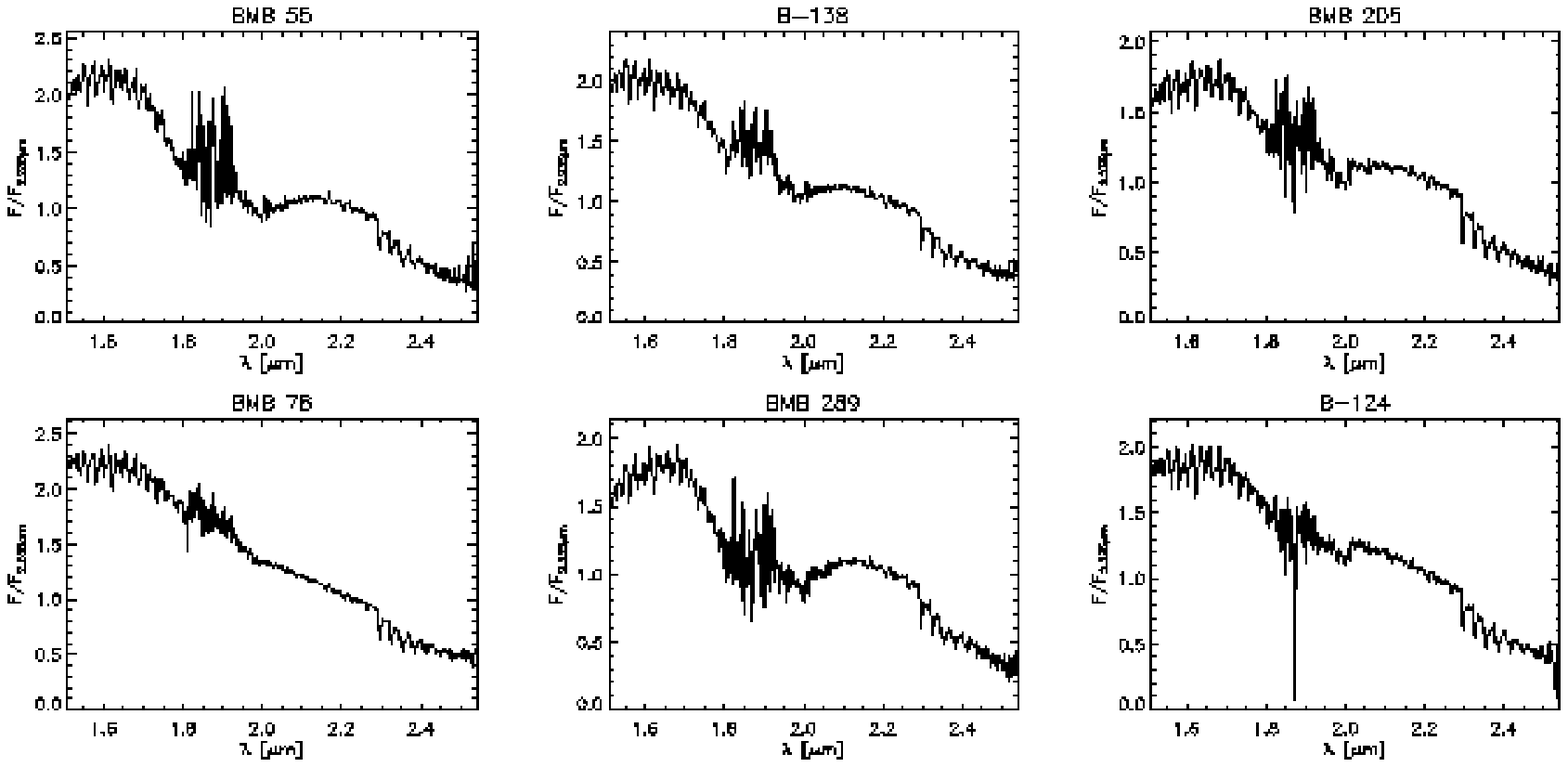}}
\caption{Stars in common with \citet{Ramirez2000}.}
\label{figure Ramirez2000}
\end{figure*}

\begin{figure*}[h]
\resizebox{\hsize}{!}{\includegraphics{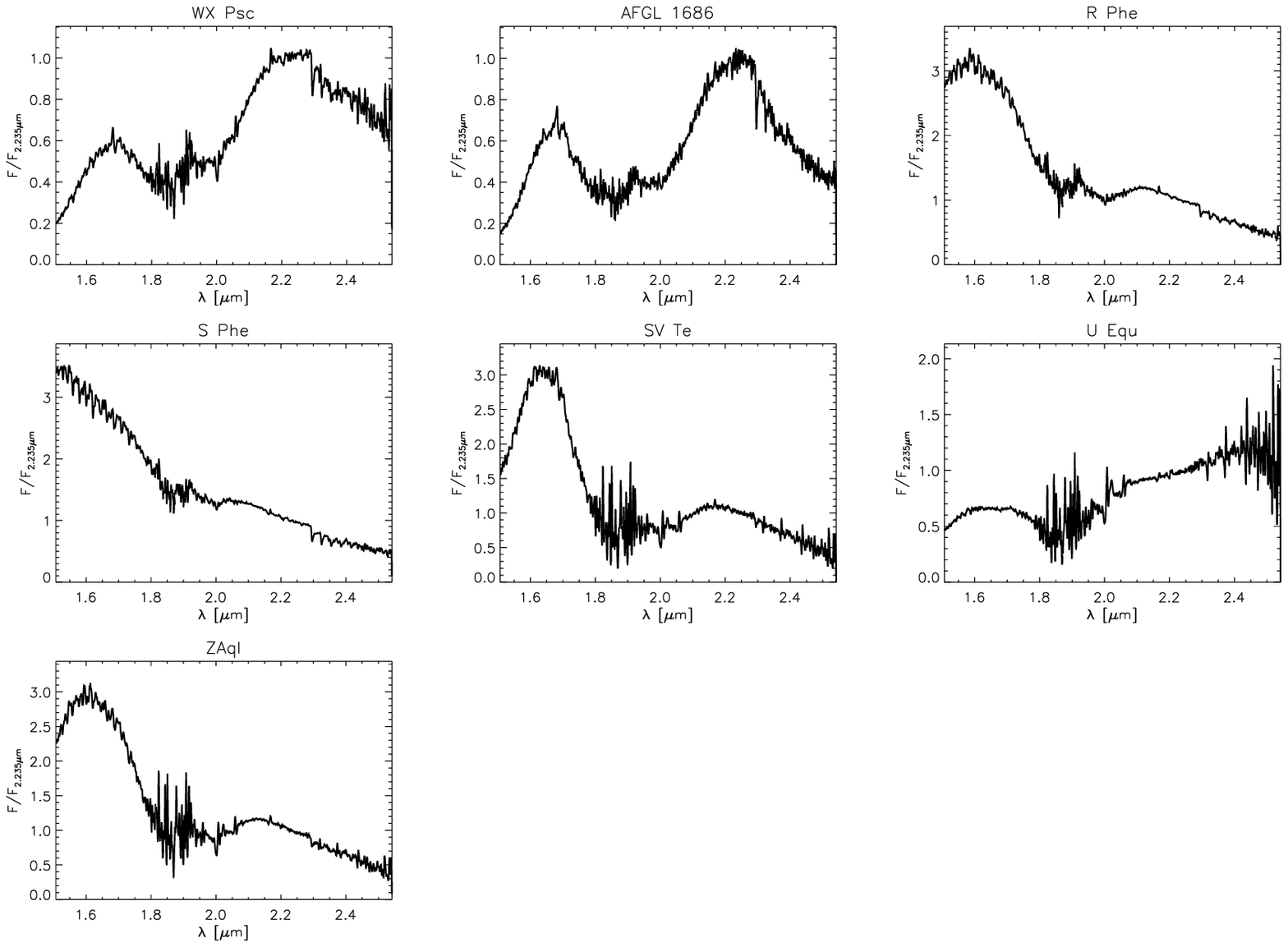}}
\caption{Stars in common with \citet{Lancon2000}.}
\label{figure Lancon and Wood}
\end{figure*}

\end{center}


\bibliographystyle{aa}
\bibliography{/home/evelien/LaTeX/bibtex/bibfile}

\end{document}

%% file: rrmacros2e.tex
\newcount\longrefs
\def\aap{\ifnum\longrefs=1 {Astron.\ Astrophys.}\else 
                           {A\hbox{\rm \&}A}\fi}
\def\aapr{\ifnum\longrefs=1 {Astron.\ Astrophys.\ Rev.}\else 
                            {A\hbox{\rm \&}AR}\fi}
\def\aaps{\ifnum\longrefs=1 {Astron.\ Astrophys.\ Suppl.}\else 
                            {A\hbox{\rm \&}AS}\fi}
\def\aj{\ifnum\longrefs=1 {Astron.\ J.}\else 
                          {AJ}\fi} 
\def\ao{\ifnum\longrefs=1 {Applied Optics}\else 
                           {Appl.\ Opt.}\fi} 
\def\aspcs{\ifnum\longrefs=1 {Astron.\ Soc.\ Pacific Conf. Series}\else 
                           {ASP Conf.\ Ser.}\fi} 
\def\apj{\ifnum\longrefs=1 {Astrophys.\ J.}\else 
                           {ApJ}\fi} 
\def\apjl{\ifnum\longrefs=1 {Astrophys.\ J. Lett.}\else 
                            {ApJ}\fi} 
\def\aplett{\ifnum\longrefs=1 {Astrophys.\ J. Lett.}\else 
                            {ApJ}\fi} 
\def\apjs{\ifnum\longrefs=1 {Astrophys.\ J. Suppl.}\else 
                            {ApJS}\fi}
\def\apss{\ifnum\longrefs=1 {Astrophys.\ and Space Science}\else 
                            {Ap\hbox{\rm \&}SS}\fi}
\def\araa{\ifnum\longrefs=1 {Ann.\ Rev.\ Astron.\ Astrophys.}\else 
                            {ARA\hbox{\rm \&}A}\fi}
\def\azh{\ifnum\longrefs=1 {Astronomicheskii Zhurnal}\else 
                            {Astron.\ Zhur.}\fi}
\def\baas{\ifnum\longrefs=1 {Bull.\ Am.\ Astron.\ Soc.}\else 
                            {BAAS}\fi}
\def\bain{\ifnum\longrefs=1 {Bull.\ Astronom.\ Institutes Netherlands}\else
                            {Bull.\ Astr.\ Inst.\ Neth.}\fi}
\def\gca{\ifnum\longrefs=1 {Geochim.\ Cosmochim.\ Acta}\else 
                           {Geochim.\ Cosmochim.\ Acta}\fi}
\def\grl{\ifnum\longrefs=1 {Geophys.\ Res.\ Lett.}\else 
                           {Geoph.\ Res.\ Lett.}\fi}
\def\iaucirc{\ifnum\longrefs=1 {IAU Circulars}\else 
                          {IAU Circ.}\fi}
\def\ip{\ifnum\longrefs=1 {in press}\else 
                          {in press}\fi}
\def\jgr{\ifnum\longrefs=1 {J.\ Geophys.\ Res.}\else 
                           {J.\ Geophys.\ Res.}\fi}  
\def\jrasc{\ifnum\longrefs=1 {J.\ Royal Astron.\ Soc.\ Canada}\else 
                           {JRAS Can.}\fi}  
\def\mnras{\ifnum\longrefs=1 {Mon.\ Not.\ Roy.\ Astron.\ Soc.}\else 
                             {MNRAS}\fi} 
\def\nat{\ifnum\longrefs=1 {Nature}\else 
                           {Nat}\fi}
\def\pasj{\ifnum\longrefs=1 {Pub.\ Astron.\ Soc.\ Japan}\else 
                            {PASJ}\fi} 
\def\pasp{\ifnum\longrefs=1 {Pub.\ Astron.\ Soc.\ Pacific}\else 
                            {PASP}\fi} 
\def\physscr{\ifnum\longrefs=1 {Physica Scripta}\else 
                            {Phys.\ Scrip.}\fi} 
\def\planss{\ifnum\longrefs=1 {Planetary \& Space Science}\else 
                            {Plan. \& Space Sci.}\fi} 
\def\procspie{\ifnum\longrefs=1 {Proc.\ SPIE}\else 
                            {Proc.\ SPIE}\fi} 
\def\qjras{\ifnum\longrefs=1 {Quarterly J.\ Royal Astron.\ Soc.}\else 
                            {QJRAS}\fi} 
\def\sa{\ifnum\longrefs=1 {Soviet Astron..}\else 
                               {Sov.\ Astron.}\fi}
\def\skytel{\ifnum\longrefs=1 {Sky \& Telescope}\else 
                            {Sky \& Tel.}\fi} 
\def\solphys{\ifnum\longrefs=1 {Solar Phys.}\else 
                               {Solar Phys.}\fi}
\def\ssr{\ifnum\longrefs=1 {Space Science Rev.}\else 
                               {Space\ Sci.\ Rev.}\fi}

\def\nl{,\ } 

\def\Leuven{Instituut voor Sterrenkunde\nl K.U.Leuven\nl Celestijnenlaan 200B\nl B--3001 Leuven\nl Belgium}

\def\Strasbourg{Observatoire Astronomique\nl Rue de L'Universite 11\nl F--67000 Strasbourg\nl France}

\def\Wien{Institut f\"ur Astronomie\nl T\"urkenschanzstrasse 17\nl A--1180 Wien\nl Austria}

\def\dutch{\def\refname{Referenties}\def\abstractname{Samenvatting}%
  \def\bibname{Bibliografie}\def\chaptername{Hoofdstuk}%
  \def\appendixname{Bijlage}\def\contentsname{Inhoudsopgave}%
  \def\listfigurename{Lijst van figuren}\def\listtablename{Lijst van tabellen}%
  \def\indexname{Index}\def\figurename{Figuur}\def\tablename{Tabel}%
  \def\partname{Deel}\def\enclname{Bijlage(n)}\def\ccname{Ter attentie van}%
  \def\headtoname{Aan}\def\headpagename{Pagina}%
  \def\today{\number\day\space\ifcase\month\or januari\or februari\or maart\or%
     april\or mei\or juni\or juli\or augustus\or september\or oktober\or%
     november\or december\fi \space\number\year}%
  \typeout{
              >>>>> use hlatex209 for Dutch hyphenation <<<<< 
         }}
\newcounter{onefig} \newcounter{fignumber}
\newcount\nocaptions \newcount\nofigures \newcount\figwidth
\newcount\viewgraphs
  \def\paper{}  \def\figlabel{} 
\long\def\nextfig#1{\setcounter{figure}{\value{fignumber}}
  \addtocounter{fignumber}{1}
  \ifnum \viewgraphs=1 \newpage \pagestyle{empty} \fi 
  \ifnum\value{onefig}=0 #1 \fi                 
  \ifnum\value{onefig}=\value{fignumber} #1 \fi}
\def\figwidths#1#2{\ifnum \nocaptions=1 #2mm \else #1mm \fi}  
\def\paper#1{}  
\long\def\plotfig#1#2{\ifnum \nofigures=1 \else #2 \fi}
\long\def\captiontext#1{\ifnum \nofigures=1 \raggedright \fi 
   \ifnum \nocaptions=1 \paper
     \ifnum \viewgraphs=0 
       \newline  \mbox{}\hrulefill\mbox{} \newline 
       \newline label:~\{\figlabel\} 
     \fi 
     \else \ifnum \nofigures=0 \fi 
   #1 \fi}

\newcount\panelwidth \newcount\panelheight 
\newcount\bxmin \newcount\bymin \newcount\bxmax \newcount\bymax
\newcount\tbxmin \newcount\tbymin
\newcount\tpanelwidth \newcount\tpanelheight \newcount\tpdif
\def\panelsize #1,#2;{\panelwidth=#1 \panelheight=#2}  
\def\setbb #1,#2;#3,#4;#5,#6;{
  \tbxmin=#1 \tbymin=#2    
  \bxmin=#3 \bymin=#4      
  \bxmax=#5 \bymax=#6}     
\def\barepanel #1{%
  \ifnum\panelheight=0 
    \tpdif=\bymax \advance\tpdif by -\bymin
    \multiply \tpdif by \panelwidth
    \tpanelheight=\tpdif
    \tpdif=\bxmax \advance\tpdif by -\bxmin
    \divide \tpanelheight by \tpdif
  \else \tpanelheight=\panelheight \fi
  \epsfig{file=#1,%
     bbllx=\bxmin bp,bblly=\bymin bp,bburx=\bxmax bp,bbury=\bymax bp,clip=,%
     width=\panelwidth mm,height=\tpanelheight mm}}
\def\labelypanel #1{
  \ifnum\panelheight=0 
    \tpdif=\bymax \advance\tpdif by -\bymin
    \multiply \tpdif by \panelwidth
    \tpanelheight=\tpdif
    \tpdif=\bxmax \advance\tpdif by -\bxmin
    \divide \tpanelheight by \tpdif
  \else \tpanelheight=\panelheight \fi
  \tpdif=\bxmax \advance\tpdif by -\tbxmin
  \tpanelwidth=\panelwidth \multiply \tpanelwidth by \tpdif
  \tpdif=\bxmax \advance\tpdif by -\bxmin
  \divide \tpanelwidth by \tpdif
  \epsfig{file=#1,%
    bbllx=\tbxmin bp,bblly=\bymin bp,bburx=\bxmax bp,bbury=\bymax bp,%
    clip=,width=\tpanelwidth mm,height=\tpanelheight mm}}
\def\labelxpanel #1{%
  \ifnum\panelheight=0 
    \tpdif=\bymax \advance\tpdif by -\bymin
    \multiply \tpdif by \panelwidth
    \tpanelheight=\tpdif
    \tpdif=\bxmax \advance\tpdif by -\bxmin
    \divide \tpanelheight by \tpdif
  \else \tpanelheight=\panelheight \fi
  \tpdif=\bymax \advance\tpdif by -\tbymin
  \multiply \tpanelheight by \tpdif
  \tpdif=\bymax \advance\tpdif by -\bymin
  \divide \tpanelheight by \tpdif
  \epsfig{file=#1,%
    bbllx=\bxmin bp,bblly=\tbymin bp,bburx=\bxmax bp,bbury=\bymax bp,%
    clip=,width=\panelwidth mm,height=\tpanelheight mm}}
\def\labelxypanel #1{%
  \ifnum\panelheight=0 
    \tpdif=\bymax \advance\tpdif by -\bymin
    \multiply \tpdif by \panelwidth
    \tpanelheight=\tpdif
    \tpdif=\bxmax \advance\tpdif by -\bxmin
    \divide \tpanelheight by \tpdif
  \else \tpanelheight=\panelheight \fi
  \tpdif=\bxmax \advance\tpdif by -\tbxmin
  \tpanelwidth=\panelwidth \multiply \tpanelwidth by \tpdif
  \tpdif=\bxmax \advance\tpdif by -\bxmin
  \divide \tpanelwidth by \tpdif 
  \tpdif=\bymax \advance\tpdif by -\tbymin 
  \multiply \tpanelheight by \tpdif
  \tpdif=\bymax \advance\tpdif by -\bymin
  \divide \tpanelheight by \tpdif
  \epsfig{file=#1,%
    bbllx=\tbxmin bp,bblly=\tbymin bp,bburx=\bxmax bp,bbury=\bymax bp,%
    clip=,width=\tpanelwidth mm,height=\tpanelheight mm}}

\def\CC{\par \vspace*{-2ex} \footnotesize \baselineskip=8pt \begin{verbatim}}

\long\def\startignore #1\stopignore{}   

\def\setlistparams{         
  \topsep=0.7ex                 
  \itemsep=0.7ex                
  \leftmargini=3ex}             
\setlistparams                  

\newcounter{alistindex}       

\newcounter{romenumnr}

\newlength{\minipagewidth}

\newsavebox{\boxcontent}
\newcommand{\ovalhead}[1]{
  \unitlength=1cm
  \sbox{\boxcontent}{\mbox{~~{#1}~~}}
  \begin{center}
    \ifdim\wd\boxcontent>6ex 
    \ifdim\wd\boxcontent<8cm 
    \begin{picture}(8,3) \thicklines     
      \put(4.0,0.8){\oval(8,1.6)} 
      \put(0.0,0.7){\parbox{8cm}{
         \begin{center} \usebox{\boxcontent} \end{center}}}
    \end{picture}
    \else \ifdim\wd\boxcontent<12cm 
    \begin{picture}(12,3) \thicklines     
        \put(6.0,0.8){\oval(12,1.6)} 
        \put(0.0,0.7){\parbox{12cm}{
           \begin{center} \usebox{\boxcontent} \end{center}}}
    \end{picture}
    \else
    \begin{picture}(16,3) \thicklines     
        \put(8.0,0.8){\oval(16,1.6)} 
        \put(0.0,0.7){\parbox{16cm}{
           \begin{center} \usebox{\boxcontent} \end{center}}}
    \end{picture}
    \fi \fi \fi
  \end{center}} 

\newcounter{headnr}            
\newcounter{subheadnr}[headnr]
\newcounter{subsubheadnr}[subheadnr]
\def\head #1\par{
  \stepcounter{headnr}                          
  \vspace{2ex} \noindent                        
  {\bf \theheadnr~~~~#1}\\[1ex] \noindent}      
\def\subhead #1\par{  
  \stepcounter{subheadnr}
  \vspace{1.3ex} \noindent
  {\bf \theheadnr.\arabic{subheadnr}~~~#1}\\[0.3ex] \noindent}
\def\subsubhead #1\par{
  \stepcounter{subsubheadnr}
  \vspace{1.0ex} \noindent
  {\bf \theheadnr.\arabic{subheadnr}.\arabic{subsubheadnr}~~~#1}\\ \noindent}

\font\dropfont= cmr12 scaled \magstep5
\def\dropcap#1#2{{\noindent
    \setbox0\hbox{\dropfont #1}\setbox1\hbox{#2}\setbox2\hbox{(}%
    \count0=\ht0\advance\count0 by\dp0\count1\baselineskip
    \advance\count0 by-\ht1\advance\count0by\ht2
    \dimen1=.5ex\advance\count0by\dimen1\divide\count0 by\count1
    \advance\count0 by1\dimen0\wd0
    \advance\dimen0 by.25em\dimen1=\ht0\advance\dimen1 by-\ht1
    \global\hangindent\dimen0\global\hangafter-\count0
    \hskip-\dimen0\setbox0\hbox to\dimen0{\raise-\dimen1\box0\hss}%
    \dp0=0in\ht0=0in\box0}#2}

\def\HtwoO{\mbox{H$_2$O}}        

\def\level #1 #2#3#4{$#1 \: ^{#2} \mbox{#3} ^{#4}$}   

\def\micron{\hbox{$\mu$m}}

\def\Teff{\hbox{$\rm{T}_{\rm eff}$}}            

\def\Msun{\hbox{M$_{\odot}$}}               

\def\mathstacksym#1#2#3#4#5{\def#1{\mathrel{\hbox to 0pt{\lower 
    #5\hbox{#3}\hss} \raise #4\hbox{#2}}}}

\mathstacksym\lta{$<$}{$\sim$}{1.5pt}{3.5pt} 
\mathstacksym\gta{$>$}{$\sim$}{1.5pt}{3.5pt} 
\mathstacksym\lrarrow{$\leftarrow$}{$\rightarrow$}{2pt}{1pt} 
\mathstacksym\lessgreat{$>$}{$<$}{3pt}{3pt} 
